\documentclass[11pt,a4paper]{article}
\usepackage{amsmath}
\usepackage[mathcal]{eucal}
\usepackage{latexsym}
\usepackage{amssymb}
\begin{titlepage}
\title{Generalized Kodama partition functions: A preview into normalizability for the generalized Kodama states.}
\author{Eyo Eyo Ita III}
\input amssym.def
\input amssym.tex

\begin{document}
\maketitle
\medskip

\begin{abstract}
\par
\medskip
\indent
In this paper we outline the computation of the partition function for the generalized Kodama states (GKod) of quantum gravity using the background field method.  We show that the coupling constant for GKod is the same dimensionless coupling constant that appears in the partition function of the pure Kodama state (Chern--Simons functional) and argue that the GKod partition function is renormalizable as a loop expansion in direct analogy to Chern--Simons perturbation theory.  The GKod partition function contains an infinite set of 1PI vertices uniquely fixed, as a result of the semiclassical-quantum correspondence, by the first-order vertex.  This implies the existence of a well-defined effective action for the partition function since the `phase' of the GKod, provided a finite state exists, is equivalent to this effective action.  Additionally, the separation of the matter from the gravitational contributions bears a resemblance to the infinite dimensional analogue to Kaluza--Klein theory.  Future directions of research include extension of the computations of this paper to the norm of the GKod as well as to examine the analogue of the Chern--Simons Jone's polynomials and link invariants using the GKod as a measure.

\end{abstract}
\end{titlepage}

\section{Introduction}
\par
\medskip
\indent 
The purpose of this paper is to apply a special technique for computing partition functions to the computation of the norm of the generalized Kodama states of quantum gravity introduced in \cite{EYO}.  The closest analogy is the partition function for the Chern--Simons functional.  There exist in the literature various methods for computing the partition function of the Chern--Simons wavefunction (\cite{PERT},\cite{PERT1},\cite{PERT3} and refences therein).  As a path integral, the nonabelian Chern--Simons theory in three spatial to the present author's knowledge dimensions produces a renormalizable loop expansion, since any quantum divergences can always be absorbed, as shown by various authors, into a redefinition of tree-level coupling constants.  This implies that the classical action and the effective action for the Chern--Simons functional are the same, to within re-definition of these constants.\par
\indent
The first attempt to apply the perturbation theory of the Chern--Simons partition function to the fill theory of quantum gravity in the connection representation appears to be \cite{CSPT}, in which Chopin Soo computes the partition function for the pure Kodama state.  We would like to generalize the techniques for computation of the Chern--Simons partition function from the pure Kodama 
state $\Psi_{Kod}$ to the generalized Kodama states $\Psi_{GKod}$.  In the present work we will apply the background field method for computing the effective action, usually attributed to Wilson and DeWitt, for the task as set out.\par
\indent  
We would ultimately like to extend the results of this paper to address the normalizability 
of $\Psi_{GKod}$.  This will depend upon the ability to make sense of the configuration path integral, for the partition function, which will in turn depend upon the finiteness of the generalized Kodama state.  We will not focus in this paper on such issues as gauge fixing and the renormalization group theory.  The main thrust is to show that the perturbative expansion exists and conforms to the requirements of a perturbatively renormalizable effective action, by direct analogy to the case for Chern--Simons theory which has been thorougly explored by other authors.  Our primary contribution is that the principle of the semiclassical-quantum correspondence introduced in \cite{EYO} is particularly fortuitous when applied particularly to this special class of quantum states $\Psi_{GKod}$.  We also present a first Ansatz for the norm of $\Psi_{GKod}$ as well as expectation values of operators on the state, not commenting in detail on reality conditions.\par
\indent
An important aspect of this paper is that all quantities are computed on the spatial 
hypersurface $\Sigma_T$ on which which $\Psi_{GKod}$ is evaluated.  So while it may appear that the treatment is timeless, we take time to be an implicit label.  We will often in this paper not distinguish between spacetime 
position $x=(\boldsymbol{x},t)$ and spatial position $\boldsymbol{x}$.  The partition function can be seen as a unified gravity-matter system with a dimensionless coupling constant.  When split into its individual components, the partition function has an interesting geometric interpretation bearing a distinct resemblance to Kaluza--Klein like theories on the infinite dimensional space of fields.  We will argue for renormalizability and finiteness of the generalized Kodama partition function in direct analogy to that for the partition function for the pure Kodama state $\Psi_{Kod}$.\par
\indent  
The format of this paper is as follows.  In section 2 we illustrate in detail the computation of the Chern--Simons partition function via the background field method, to set the stage and to cast the formalism into the language of wavefunctions.  In section 3 we motivate the notion of the existence of a wavefunction of the universe whose norm should be computed.  In section 4 we outline the computation of some one particle-irreducible (1PI) vertices of $\Psi_{GKod}$ highlighting their geometric interpretation in analogy to Kaluza--Klein theory on infinite dimensional spaces.  In section 5 we compute the generalized Kodama partition function, taking into account the Gaussian and all higher order contributions, from these 1PI vertices.  In section 6 we write down an expression for the generalized Kodama partition function both perturbatively and as well in terms of an interesting nonperturbative representation\footnote{This requires some concepts from quantum field theories on curved spacetimes.} obtaining a loop expansion in the dimensionless constant $\sqrt{\hbar{G}\Lambda}$.  In section 7 we rewrite the quantum Hamiltonian constraint of \cite{EYO} in the language of 1PI vertices, which highlights some imprints from the semicalssical limit of gravity below the Planck scale upon the generalized Kodama partition function.\par
\indent

\section{The Chern--Simons Partition Function in the language of wavefunctions}

The Chern--Simons partition function is given by the following configuration space path integral 

\begin{equation}
\label{CHERN}
Z=\int{DA}e^{-{{I[A]} \over k}}
\end{equation}

\noindent
where $k$ is a dimensionless coupling constant and $I_{CS}$ is the Chern--Simons functional on a three dimensional spatial manifold $\Sigma$ given (in the notation of differential forms) by

\begin{equation}
\label{CHERN1}
I[A]=I_{CS}[A]=\int_{\Sigma}\hbox{tr}\Bigl({A}\wedge{dA}+{2 \over 3}{A}\wedge{A}\wedge{A}\Bigr).
\end{equation}

\indent
Here $A=\tau_aA^a_idx^i$ is a $SU(2)$-valued connection and $i=1,2,3$ correspond to spatial indices in $\Sigma$.  One possible method for evaluating the configuration space integral (\ref{CHERN}) is to expand the connection in quantum fluctuations $A$ about a reference connection $\alpha$ in the 
form $A=\alpha+a$, where the reference connection $\alpha$ satisfies some predetermined criterion within the 3-manifold $\Sigma$.\footnote{In usual treatments $\alpha$ is taken to be a flat connection such that $d\alpha+{\alpha}\wedge{\alpha}=0$ and $a$ corresponds to the quantized fluctuations from the reference connection.  In the present work we will lift the restriction that $\alpha$ be flat, reserving the option to choose the reference configuration judiciously as required.}  These fluctations range over all values at each spatial point $\boldsymbol{x}$ within the manifold $\Sigma$ resulting in a loop expansion.  The connection $\alpha$ plays the role of a background field, as in the background field method \cite{BACKGROUND},\cite{JACKIW}.\par
\indent
If one defines a `reference' wavefunctional $\Psi=\Psi[\alpha]$ based upon the reference configuration corresponding to the background field $\alpha$, e.g. $\Psi[\alpha]=\hbox{exp}(-k^{-1}I[\alpha])$ then one can analyse the quantum fluctuations relative to $\Psi[\alpha]$ by implementing the path integral as a kind of transformation of this wavefunctional.  So (\ref{CHERN}) becomes 

\begin{eqnarray}
\label{CHERN2}
\Psi_{quant.}[\alpha]=\int{Da}~\hbox{exp}(-k^{-1}I_{CS}[\alpha+a])\nonumber\\
=\int{Da}~\hbox{exp}\Bigl[\int_{\Sigma}d^3{x}~{a(x)}{\delta \over {\delta\alpha(x)}}\Bigr].
\hbox{exp}(-k^{-1}I_{CS}[\alpha])
\end{eqnarray}

\noindent
The result of evaluating the path integral can be expressed in terms of loop corrections relative to the reference wavefunctional via the relation
\begin{equation}
\label{CHERN3}
\hat{\O}\Psi[\alpha]=\Bigl(\sum_{n=0}^{\infty}L_n[\alpha]k^{n/2}\Bigr)\Psi[\alpha]
\end{equation}

\noindent
where we have defined in (\ref{CHERN3}) a path integration operator $\hat{\O}$, given by

\begin{equation}
\label{CHERN4}
\hat{\O}=\int{Da}~\hbox{exp}\Bigl[\int_{\Sigma}d^3{x}~{a(x)}{\delta \over {\delta\alpha(x)}}\Bigr]
=\int{Da}~\hbox{exp}\Bigl[{1 \over k}\int_{\Sigma}d^3{x}{a(x)}\hat\pi(x)\Bigr].
\end{equation}

\noindent
and $L_n=L_n[\alpha]$ is the sum of all Feynman diagrams diagrams for 
the $n^{th}$ loop with appropriate symmetry factors, which depends explicitly upon the reference configuration $\alpha$ chosen. Equation (\ref{CHERN3}) is the loop expansion about the 
configuration $\alpha=\alpha(\boldsymbol{x}$.  From the functional translation operator 
in (\ref{CHERN4}) One can associate a conjugate momentum operator 
$\hat\pi$ to the reference connection $\alpha$, which satisfies the equal-time commutation relations\footnote{By definition the times are equal since we are evaluating the partition function with respect to the hypersurface $\Sigma_T$ upon which the wavefunction is defined.}

\begin{eqnarray}
\label{COMMU}
\bigl[\hat{\alpha}^a_i(\boldsymbol{x},t),\hat{\pi}^j_b(\boldsymbol{y},t)\bigr]
=k\delta^a_b\delta^j_i\delta^{(3)}(\boldsymbol{x}-\boldsymbol{y})
\end{eqnarray}

\noindent
given in the Schr\"odinger representation by

\begin{equation}
\label{CHERN5}
\hat\pi(x)\Psi[\alpha]=k{{\delta} \over {\delta\alpha(x)}}\Psi[\alpha].
\end{equation}

\indent
One can then evaluate the multiple functional derivatives required for (\ref{CHERN2}), which correspond to vertices in a Feynman diagrammatic expansion.  Let us evaluate these explicitly.  The zeroth order vertex is just $I_{CS}[\alpha]$.  The first-order (1-point) vertex is

\begin{equation}
\label{CHERN6}
{{\delta{I_{CS}}} \over {\delta{A^a_{i}(x)}}}[A]\biggl\vert_{A=\alpha}=B^i_{a}[\alpha(x)]
=\epsilon^{ijk}\bigl[\delta_{ae}\partial_{j}\alpha^e_{k}(x)
+{1 \over 2}f_{aed}\alpha^e_{j}(x)\alpha^d_{k}(x)\bigr],
\end{equation}

\noindent
and the second-order (2-point) vertex is given by

\begin{eqnarray}
\label{CHERN7}
{{\delta^2I[A]} \over {\delta{A^a_{i}(x)}\delta{A^b_{j}(y)}}}\equiv{D^{ij}_{ab}(\boldsymbol{x},\boldsymbol{y})}\biggl\vert_{A=\alpha}
={\delta \over {\delta{A^b_{j}(y)}}}B^i_{a}(x)\biggl\vert_{A=\alpha}\nonumber\\
={\delta \over {\delta{A^b_{j}(y)}}}\epsilon^{ijk}\bigl[\delta_{ae}\partial_{j}A^e_{k}(x)+{1 \over 2}f_{aed}A^e_{j}(x)A^d_{k}(x)\bigr]\biggl\vert_{A=\alpha}\nonumber\\
=\epsilon^{ijk}\bigl[\delta_{ab}\partial_{k}+f_{abe}\alpha^e_{k}(x)\bigr]
\delta^{(3)}(\boldsymbol{x}-\boldsymbol{y}).
\end{eqnarray}

\noindent
The object (\ref{CHERN7}) can be thought of as the kinetic operator for Chern-Simons theory in a particular gauge.  For a trivial connection it would correspond to a position space propagator

\begin{equation}
\label{CHERN8}
(\Delta_0)^{ab}_{ij}(\boldsymbol{x},\boldsymbol{y})=
[D_0^{-1}(x,y)]^{ab}_{ij}\equiv\delta^{ab}\epsilon_{ijk}
{{x^k-y^k} \over {\vert\boldsymbol{x}-\boldsymbol{y}\vert^3}}.
\end{equation}

\noindent
where we have define the `bare' propagator as $\Delta_0$.  However, when twisted by a 
connection $\alpha$ it generalize the Coulomb Green's function to $SU(2)$ nonabelian gauge theory in analogy to \cite{SCHROREP}

\begin{eqnarray}
\label{CHERN9}
[D^{-1}(x,y)]^{ab}_{ij}
=\delta^{ab}\epsilon_{ijk}
{{x^k-y^k} \over {\vert\boldsymbol{x}-\boldsymbol{y}\vert^3}}
+\sum_{n=1}^{\infty}(-g)^n\int{d^3x_1}\dots\int{d^3x_n}
\alpha_{i_1}^{a_1}\alpha_{i_2}^{a_2}\dots\alpha_{i_n}^{a_n}
\nonumber\\
\times{1 \over {\vert{x-x_1}\vert}}\partial_{x_1}^{a_1}
{1 \over {\vert{x_1-x_2}\vert}}\partial_{x_2}^{a_2}
{1 \over {\vert{x_2-x_3}\vert}}\partial_{x_3}^{a_3}
\dots
{1 \over {\vert{x_n-y}\vert}}\partial_{x_n}^{a_n}
\end{eqnarray}

\noindent
where we have included the coupling constant $g$ in the deefinition of the $SU(2)$ covariant derivative.\footnote{The `dressed' propagator for a general theory can in general be computed from its pure version by heat kernel methods as in \cite{AVRA2},\cite{AVRA3},\cite{AVRA4}.}

\par
\indent
The third-order (3-point) vertex is given by

\begin{eqnarray}
\label{CHERN10}
{{\delta^3I_{CS}[A]} \over {\delta{A^a_{i}(x)\delta{A^b_{j}(y)\delta{A^c_{k}(z)}}}}}
={\delta \over {\delta{A^c_{k}(z)}}}{D^{ij}_{ab}(\boldsymbol{x},\boldsymbol{y})}
=\epsilon^{ijk}_{abc}(\boldsymbol{x},\boldsymbol{y},\boldsymbol{z})\nonumber\\
=\epsilon^{ijk}f_{abc}\delta^{(3)}(\boldsymbol{x}-\boldsymbol{y})\delta^{(3)}(\boldsymbol{x}-\boldsymbol{z}).
\end{eqnarray}

\noindent
As all occurences of $A^a_i$ in (\ref{CHERN10}) have been exhausted, the fourth and higher functional derivatives vanish and we find that there are two structures that can appear
in a loop expansion of this partition function, namely the propagators, labelled by $D^{ij}_{ab}$ and the trivalent vertices labelled by 
$\epsilon^{ijk}f_{abc}$.\footnote{We will find that in the partition function corresponding to the generalized Kodama states $\Psi_{GKod}$, additional structures appear due to the
presence of quantized matter fields interacting with quantized gravity.}  So the functional Taylor expansion of the Chern-Simons functional about the reference connection $\alpha$ is given by  

\begin{eqnarray}
\label{CHERN11}
I_{CS}[\alpha+a]=I_{CS}[\alpha]
+\int{d^3{x}}a^a_{i}(x)B^i_{a}(x)
+{1 \over {2!}}\int{d^3{x}}\int{d^3{y}}a^a_{i}(x)D^{ij}_{ab}(\boldsymbol{x},\boldsymbol{y})a^b_{j}(y)\nonumber\\
+{1 \over {3!}}\int{d^3{x}}\int{d^3{y}}\int{d^3{z}}\epsilon^{ijk}f_{abc}a^a_{i}(x)a^b_{j}(y)a^c_{k}(z)\delta^{(3)}(\boldsymbol{x}-\boldsymbol{y})\delta^{(3)}(\boldsymbol{x}-\boldsymbol{z})
\end{eqnarray}

\noindent
where $D^{ij}_{ab}$ is given by

\begin{equation}
\label{CHERN12}
D^{ij}_{ab}(\boldsymbol{x},\boldsymbol{y})=\epsilon^{ijk}\bigl[\delta_{ab}\partial_{k}+f_{abe}{\alpha}^e_{k}(x)\bigr]\delta^{(3)}(\boldsymbol{x}-\boldsymbol{y}).
\end{equation}

\noindent
Note that the functional terminates at cubic order.\footnote{One should expect the effective action as well to terminate at cubic order since Chern--Simons theory is perturbatively renormalizable.}
It is convenient to express this in shorthand notation if we associate indices (labels) that are naturally grouped together, e.g.

\begin{eqnarray}
\label{CHERN13}
i\equiv{(i,a,\boldsymbol{x})}~~;~~j\equiv{(j,b,\boldsymbol{y})}~~;k\equiv{(k,c,\boldsymbol{z})}\nonumber\\
D^{ij}_{ab}(\boldsymbol{x},\boldsymbol{y})\longrightarrow{D^{ij}}~~;~~\epsilon^{ijk}_{abc}(\boldsymbol{x},\boldsymbol{y},\boldsymbol{z})\longrightarrow\varepsilon^{ijk}
\end{eqnarray}

\noindent
and contract over all indices, including integration.  We then have

\begin{equation}
\label{CHERN14}
I_{CS}[\alpha+a]=I_{CS}[\alpha]+a_{i}B^i+{1 \over {2}}a_{i}D^{ij}a_j
+{1 \over {6}}\varepsilon^{ijk}a_{i}a_{j}a_k
\end{equation}

\noindent
Expression of infinite dimensional spaces in shorthand notation in terms of finite dimensional ones facilitates visualization of the loop expansion in terms of Feynman diagrams.  By completion of the square,

\begin{equation}
\label{CHERN15}
I_{CS}[\alpha+a]=I_0+{1 \over 2}(a+BD^{-1})_{i}D^{ij}(a+BD^{-1})_{j}
-{1 \over 2}B^i(D^{-1})_{ij}B^j+{1 \over {6}}\varepsilon^{ijk}a_{i}a_{j}a_k.
\end{equation}

\noindent
To evaluate the path integral one makes a shift $a\rightarrow{a-BD^{-1}}$ yielding

\begin{eqnarray}
\label{CHERN16}
I_{CS}[\alpha+a-BD^{-1}]=I_0+{1 \over 2}a_{i}D^{ij}a_{j}-{1 \over 2}B^i(D^{-1})_{ij}B^j\nonumber\\
+{1 \over 6}\varepsilon^{ijk}(a-BD^{-1})_{i}(a-BD^{-1})_{j}(a-BD^{-1})_k.
\end{eqnarray}

\noindent
So the path integral to be performed is given by

\begin{eqnarray}
\label{CHERN17}
\Psi_{eff}[\alpha]=\Psi[\alpha]\hbox{exp}\Bigl[-{1 \over {2k}}B^i(D^{-1})_{ij}B^j\Bigr]
\int{Da}\hbox{exp}\Bigl[-{1 \over {2k}}a_{i}D^{ij}a_{j}\Bigr]\nonumber\\
\hbox{exp}\Bigl[-{1 \over {6k}}\varepsilon^{ijk}(a-BD^{-1})_{i}(a-BD^{-1})_{j}(a-BD^{-1})_k\Bigr]
\end{eqnarray}

\noindent
Let us now interpret the factors 
in (\ref{CHERN17}), as we will be performing the analogue of this procedure for the generalized Kodama wavefunction including matter.  The term $\Psi[\alpha]=e^{-1/k(I[\alpha])}$ can be thought of as the reference wavefunction without any quantum corrections, which is the Chern-Simons functional evaluated on the reference connection.\par
\indent  
The second factor of (\ref{CHERN17}) in the usual treatments of Chern-Simons perturbation theory would usually be 1, if the Chern-Simons partition function is about a flat 
connection $B^i_{a}(x)=0~~\forall{x}$.  However, we reserve the freedom to choose the 
configuration $B^i_a$ appropriately, which in quantum gravity will correspond to the Ashtekar curvature.\footnote{This so as to encompass the interaction of gravity coupled to matter fields quantized on the same footing.} 

\begin{equation}
\label{CHERN19}
\hbox{exp}\Bigl[{1 \over {2k}}B^i(D^{-1})_{ij}B^j\Bigr]
=\hbox{exp}\Bigl[{1 \over {2k}}\iint{d^3{x}}{d^3{y}}B^i_{a}(\boldsymbol{x})
(D^{-1}(\boldsymbol{x},\boldsymbol{y}))^{ab}_{ij}B^j_{b}(\boldsymbol{y})\Bigr].
\end{equation}

\noindent
Equation (\ref{CHERN19}) has an interesting interpretation as the analogue of the ground state wavefunction for Maxwell theory in the Schr\"odinger representation, given by \cite{WHEELER}

\begin{equation}
\label{CHERN20}
\Psi_0[A]=\hbox{exp}\Bigl[{1 \over {2}}\iint{d^3{x}}
{d^3{y}}B^{i}(x){1 \over {\vert\boldsymbol{x}-\boldsymbol{y}\vert^2}}B_{i}(y)\Bigr]
\end{equation}

\noindent
This corresponds to the starting Lagrangian of a nonlocal curvature-squared theory in three dimensions.\footnote{A subtlety of (\ref{CHERN19}) and (\ref{CHERN20}) is the label $x=(\boldsymbol{x},t)$ in the magnetic field $B^i_a$ as compared with the bold version $\boldsymbol{x}$ occuring in the propagator.  This signifies that the propagation occurs within a fixed spatial hypersurface and not in the time direction, as with the instantaneous Coulomb interaction.}  The final two terms of (\ref{CHERN17}) can be evaluated in terms of a loop 
expansion either by steepest descents approximation, or by introducing source currents into the third term, e.g.

\begin{eqnarray}
\label{CHERN21}
\int{Da}\hbox{exp}\Bigl[-{1 \over {2k}}a_{i}D^{ij}a_{j}\Bigr]
\longrightarrow\int{Da}\hbox{exp}\Bigl[-{1 \over {2k}}a_{i}D^{ij}a_{j}+a_{i}J^i\Bigr]\nonumber\\
=k^{{1 \over 2}\zeta(0)}\tau[A]e^{-{{\zeta^{\prime}_{D}(0)} \over 2}}
\hbox{exp}\Bigl[{k \over 2}J^{i}(D^{-1})_{ij}J^{j}\Bigr]
\end{eqnarray}

\noindent
Here in (\ref{CHERN21}), $\tau[A]$ is the Ray-Singer torsion corresponding to the connection's Laplacian as it relates to zero-forms (ghosts) and to one-forms.  It is a numerical
constant common to all of the states which should can out in the calculation of expectation values and relative probabilities.\footnote{See \cite{ABELDUAL} for some interesting results concerning invariant integration in gauge theory}  We have for simplicity not included any contribution from ghosts and gauge-fixing, by restricting $D^{-1}$ to the orthogonal complement to the null space of of the kinetic operator $D^{ij}_{ab}$.\footnote{A good review of the structures required for gauge-fixing the path integral in Chern--Simons theory can be found in \cite{PERT2}.}\par
\indent
Then the final two terms of (\ref{CHERN17}) can then be combined into a loop expansion of the form

\begin{equation}
\label{CHERN22}
\hbox{exp}\Bigl[-{1 \over {6k}}\varepsilon^{ijk}\bigl({\delta \over {\delta{J}}}-BD^{-1}\bigr)_{i}
\bigl({\delta \over {\delta{J}}}-BD^{-1}\bigr)_{j}
\bigl({\delta \over {\delta{J}}}-BD^{-1}\bigr)_k\Bigr]
\hbox{exp}\Bigl[{k \over 2}J^{i}(D^{-1})_{ij}J^{j}\Bigr]\Bigl\vert_{J=0}.
\end{equation}

\noindent
Equation (\ref{CHERN22}) can be inverted using the infinite dimensional version of identity

\begin{eqnarray}
\label{INDENTI}
F[\partial/\partial{x}]G[x]=G[\partial/\partial{y}]F[y]e^{x\cdot{y}}\biggl\vert_{y=0}.
\end{eqnarray}

\noindent
The final result for the Chern--Simons partition function is given by

\begin{eqnarray}
\label{CHERN23}
\Psi_{eff}[\alpha]=k^{{1 \over 2}\zeta(0)}\tau[A]e^{-{{\zeta^{\prime}_{D}(0)} \over 2}}\Psi[\alpha]
\hbox{exp}\Bigl[{k \over 2}B^{i}(D^{-1})_{ij}B^{j}\Bigr]
\hbox{exp}\Bigl[{k \over 2}{\delta \over {\delta{K_i}}}(D^{-1})_{ij}{\delta \over {\delta{K_j}}}\Bigr]
\nonumber\\
\hbox{exp}
\Bigl[-k^{-1}\varepsilon^{ijk}(\sqrt{k}K-BD^{-1})_{i}(\sqrt{k}K-BD^{-1})_{j}(\sqrt{k}K-BD^{-1})_k\Bigr]\Bigl\vert_{K=0}.
\end{eqnarray}

\noindent
Equation (\ref{CHERN23}) will generate a Feynman-diagrammatic loop expansion.  Thus, the final expression for the partition function appears in the form of an eigenvalue equation

\begin{eqnarray}
\label{CHERN24}
\Psi_{eff}[\alpha]=\hat{\O}\Psi[\alpha]=\Bigl(\sum_{n=0}^{\infty}L_n[\alpha]k^{n/2}\Bigr)\Psi[\alpha].
\end{eqnarray}

\noindent
The quantity in brackets in (\ref{CHERN24}) can be thought of as the eigenvalue of the path integration operator acting on the `reference' wavefunction

\begin{equation}
\label{CHERN25}
\prod_{x}\int{da(x)}\hbox{exp}\Bigl[{1 \over k}\int_{\Sigma}d^3{x}~{a(x)}\hat\pi(x)\Bigr]
\sim\sum_{n=0}^{\infty}L_nk^{n/2}.
\end{equation}

\noindent
On a final note, we point out that the result of the path integration operator should be independent of the reference configuration $\alpha$ when carried out to all orders in the loop expansion.  We are now ready to apply the below formalism to the computation of the norm of the generalized Kodama states $\Psi_{GKod}$ of quantum gravity.\par
\indent

\subsection{Prelude into quantum gravity}

It is argued in \cite{EYOPATH} that the equivalence of the canonical to path integration approaches to quantization of general relativity holographically determines the evolution of the wavefunction of the universe from an initial spatial hypersurface $\Sigma_0$ to a final spatial hypersurface 
$\Sigma_T$ in terms of the fields living on these hypersurfaces.  It is also argued that the wavefunction subject to the semiclassical-quantum correspondence can be written in the form 
$\Psi_{GKod}=e^{I_{GKod}}$, where the `phase' $I_{GKod}$ is fixed by the vanishing of all singularities arising from the quantum Hamiltonian constraint.  If the canonical and path integration procedures are indeed equivalent, then the wavefunction $\Psi_{GKod}$ must be the same as the path integral of the starting action for general relativity in Ashtekar variables.  This implies that one may make the identification 
$I_{GKod}\sim{I}_{eff}$, where $I_{eff}=e^{i\Gamma_{eff}}$ is the effective action of the quantum theory in terms of 1PI vertices.  The effective action for a general quantum field theory can be expressed by the four-dimensional integral of an effective Lagrangian density

\begin{equation}
\label{NORM}
I_{eff}=\int^T_{t_0}dt^{\prime}\int_{\Sigma}d^3{x}~L_{eff}(x,t^{\prime})=\int_{M}L_{eff}
\end{equation}

\noindent
However, since $I_{GKod}$ is defined on a three dimensional hypersurface $\Sigma_T$ it follows that the four dimensional version is the spacetime integral of a total time derivative, which has been shown in \cite{EYO}.  In order to compute the partition function for the generalized Kodama
 state $\Psi_{GKod}$, and utlimately its norm, two conditions are required: First, the 
wavefunction $\Psi_{GKod}$ must be finite.  Secondly, the covariant four dimensional action $I_{eff}$ in (\ref{NORM}) must be projectable in an invariant manner onto a three dimensional spatial hypersurface $\Sigma$ to define a wavefunction for the universe.  Since the history of the fields making up the four-dimensional Lagrangian density is in general freely specifiable in a quantum field theory, then it is not in general possible to deterministically write the Lagrangian density as a function of time except in the semiclassical limit.  In this way, one typically expresses the wavefunction in terms of initial and final data by expanding the path integral in quantum fluctuations about a classical solution containing that data.  In an abuse of notation, one may interchange the order of spacetime integration in (\ref{NORM})\footnote{Since all histories in the Feynman sum over histories are equally weighted} to obtain

\begin{equation}
\label{NORM1}
I_{\Sigma}=\int_{\Sigma}d^3{x}\int^T_{t_0}dt^{\prime}L_{eff}(x,t^{\prime})
=\int_{\Sigma}d^3{x}L_{\Sigma}(x).
\end{equation}

\noindent
We argue that the function $I_{\Sigma}$ is the quantity on the three-dimensional spatial boundary of spacetime which one associates to the wavefunction of the universe derived by path integral, and is the analogue of $I_{GKod}$ in quantum gravity.

\subsection{A few basic conventions and identities}

Prior to proceeding with the generalized Kodama partition function, we list for completeness the vertices corresponding to the Feynman rules for ordinary Chern-Simons perturbation theory.  In the case of quantum gravity coupled to matter fields, there will be a model-specific modification to these vertices.  For notation, indices from the beginning of the Latin alphabet $a,b,c,\dots$ correspond in quantum gravity to left-handed $SU(2)_{-}$ indices while those from the middle of the alphabet $i,j,k,\dots$ correspond to spatial indices in $\Sigma$.\par
\indent  
The Ashtekar magnetic field is given by

\begin{equation}
B^k_{c}(x)=\epsilon^{klm}\Bigl[{{\partial{A^c_m(x)}} \over {\partial{x}^l}}+{1 \over 2}f^{cde}A^d_{l}(x)A^e_m(x)\Bigr]
\end{equation}

\noindent
for left-handed $SU(2)_{-}$ structure constants $f^{abc}$.  We have the following additional identities

\begin{eqnarray}
{{\delta{B^k_c}(\boldsymbol{x},t)} \over {\delta{A^a_i}(\boldsymbol{y},t)}}
={\delta \over {\delta{A^b_j}(\boldsymbol{y},t)}}\Bigl[\epsilon^{klm}D_{l}A^c_m(\boldsymbol{x},t)\Bigr]\nonumber\\
=\epsilon^{klm}{\delta \over {\delta{A^a_i}(\boldsymbol{y},t)}}
\Bigl[{\partial \over {\partial{x^l}}}A^c_m(\boldsymbol{x},t)
+{1 \over 2}f^{cde}A^d_{l}(\boldsymbol{x},t)A^e_m(\boldsymbol{x},t)\Bigr]\nonumber\\
=\epsilon^{ikl}[\delta_{ac}\partial_l+f_{ace}A^e_l(\boldsymbol{x},t)\Bigr]\delta^{(3)}(\boldsymbol{x}-\boldsymbol{y})
\equiv\delta^{(3)}(\boldsymbol{x}-\boldsymbol{y}){D}^{ik}_{ac}.
\end{eqnarray}

\noindent
Note that in evaluating functional derivatives we are restricted to performing operations on quantities evaluated at the same time $t$, e.g. on the same spatial 3-surface $\Sigma_t$ in order to make use of the equal-time quantum commutation relations.  In analogy to Klein-Gordon theory in Minkowski spacetime for instance, the commutation relations are defined at equal times $t_1=t_2$.

\begin{equation}
[\widetilde{\sigma}^i_a(\boldsymbol{x},t_1),A^b_j(\boldsymbol{y},t_2)]=\hbar{G}\delta^i_{j}\delta^b_a\delta^{(3)}(\boldsymbol{x}-\boldsymbol{y})\longrightarrow
\hat{\widetilde{\sigma}^i_a}(\boldsymbol{x},t)=\hbar{G}{\delta \over {\delta{A}^a_i(\boldsymbol{x},t)}}
\label{CT1}
\end{equation}

\noindent
For $t_1\neq{t_2}$ eq(\ref{CT1}) does not hold and propagation of the field $A^a_i$ between these two times must be taken into account.  By restricting the variables to the same 3-surface 
$\Sigma_t$ as required for the computation of norms, we avoid the issue of time.\par
\indent  
We also have, for the second functional derivative

\begin{eqnarray}
{{\delta^2{B^k_c}(\boldsymbol{z},t)} \over {\delta{A^a_i}(\boldsymbol{x},t)\delta{A^b_j}
(\boldsymbol{y},t)}}
=\epsilon^{ijk}f_{abc}\delta^{(3)}(\boldsymbol{z}-\boldsymbol{x})\delta^{(3)}(\boldsymbol{z}-\boldsymbol{y})\nonumber\\
\equiv\epsilon^{ijk}_{abc}\delta^{(3)}(\boldsymbol{z}-\boldsymbol{x})\delta^{(3)}(\boldsymbol{z}-\boldsymbol{y}).
\end{eqnarray}

\noindent
All higher functional derivatives vanish, since $\epsilon^{ijk}_{abc}$ is a numerical constant.  From now on we will suppress the position dependence of the variables when convenient.  Finally, since the Ashtekar variables are complex, we require that the wavefunction of the 
universe $\Psi_{GKod}=\Psi_{GKod}[A^a_i,\phi^{\alpha}]$, where $\phi^{\alpha}$ represent the matter fields, be holomorphic in its dependence upon the Ashtekar connection $A^a_i$.

\section{Generalized Kodama states as the wavefunction of the universe}

If there exists such a notion as the wavefunction of the universe $\Psi_{GKod}$, then it should be possible in principle to compute its norm.  We will assume that the wavefunction can be written in the form

\begin{equation}
\label{ANZ}
\Psi_{GKod}[A,\phi]=e^{(\hbar{G})^{-1}\Gamma_{\Sigma}[A,\phi]}, 
\end{equation}

\noindent
where $\Gamma_{\Sigma}$ is a three-dimensional Lagrangian action defined on a spatial 
three-surface $\Sigma$.  Due to the equivalence of the canonical and path integration approaches to quantization \cite{WAVE3}, this wavefunction (\ref{ANZ}) has the interpretation of the starting action of quantum gravity evaluated on the solution to the quantum constraints and projected holographically to the final spatial hypersurace $\Sigma$.\footnote{Due to the choice of $\Psi_{GKod}$ as the integral of a total time derivative, as shown in \cite{EYO}.}  $\Gamma_{\Sigma}$ can then be seen as the projection of the effective action $\Gamma_{eff}$ onto a spatial hypersurface.  The quantized CDJ Ansatz on this wavefuntion (\ref{ANZ}), one of the defining relations for $\Psi_{GKod}$, is

\begin{equation}
\label{NORM1}
\hbar{G}{\delta \over {\delta{A}^a_i(x)}}\Psi_{GKod}[A,\phi]=(\Psi_{ae}(x)B^i_e(x))\Psi_{GKod}[A,\phi]
\end{equation}

\noindent
Substituting (\ref{ANZ}) into (\ref{NORM1}), we obtain the relation

\begin{equation}
\label{ANZ1}
\hbar{G}{\delta \over {\delta{A}^a_i(x)}}\Psi_{GKod}[A,\phi]
=\Bigl({{\delta\Gamma_{\Sigma}} \over {\delta{A}^a_i(x)}}\Bigr)\Psi_{GKod}[A,\phi].
\end{equation}

\noindent
Evaluating the analogous vertex for the matter field $\phi^{\alpha}$, we have

\begin{eqnarray}
\label{ANZ2}
-i\hbar{\delta \over {\delta\phi^{\alpha}(x)}}\Psi_{GKod}[A,\phi]\nonumber\\
=\Bigl({{-i} \over G}{{\delta\Gamma_{\Sigma}} \over {\delta\phi^{\alpha}(x)}}\Bigr)\Psi_{GKod}[A,\phi]=\pi_{\alpha}[A,\phi]\Psi_{GKod}[A,\phi]
\end{eqnarray}

\noindent
From (\ref{NORM1}),(\ref{ANZ1}) and (\ref{ANZ2}) one can read off the relation for the one-point vertex in terms of the gravitationally determined quantities

\begin{equation}
\label{ANZ3}
{{\delta\Gamma_{\Sigma}} \over {\delta{A}^a_i(x)}}=\Psi_{ae}(x)B^i_e(x);~~{{\delta\Gamma_{\Sigma}} \over {\delta\phi^{\alpha}(x)}}=iG\pi_{\alpha}(x).
\end{equation}

\noindent
The mixed partials condition is a statement of the triviality of the equal time commutation relations between the matter and gravitational conjugate momenta \cite{EYO}

\begin{eqnarray}
\label{PARTIALS}
\bigl[\widetilde{\sigma}^i_{a}(x),\pi_{\alpha}(y)\bigr]=\bigl[\widetilde{\sigma}^i_{a}(y),\pi_{\alpha}(x)\bigr]=0.
\end{eqnarray}

\noindent
In the language of wavefunctions (\ref{PARTIALS}) reads, making the identification in dimensionless units $I_{\Sigma}=(\hbar{G})^{-1}\Gamma_{\Sigma}$,

\begin{eqnarray}
\label{PARTIALS1}
\hbar{G}{{\delta{I_\Sigma}} \over {\delta{A}^a_{i}(x)}}=\Psi_{ae}(x)B^i_{e}(x);~~-i\hbar{{\delta{I_{\Sigma}}} \over {\delta\phi^{\alpha}(x)}}=\pi_{\alpha}(x);\nonumber\\
-i\hbar^2{G}{{\delta^2\Psi_{GKod}} \over {\delta\phi^{\alpha}(y){\delta{A}^a_i(x)}}}
=-i\hbar^2{G}{{\delta^2\Psi_{GKod}} \over {{\delta{A}^a_i(x)}\delta\phi^{\alpha}(y)}}
\end{eqnarray}

\noindent
which enables one as a consistency condition to mix partial derivatives on the conjugate momenta

\begin{eqnarray}
\label{PARTIALS2}
-i\hbar{{\delta\Psi_{ae}(x)} \over {\delta\phi^{\alpha}(y)}}B^i_e(x)=\hbar{G}{{\delta\pi_{\alpha}(x)} \over {\delta{A}^a_{i}(y)}}\longrightarrow\nonumber\\
{{\partial\pi_{\alpha}} \over {\partial{A}^a_{i}}}\delta^{(3)}(\boldsymbol{x}-\boldsymbol{y})={{-i} \over G}B^i_{e}{{\partial\Psi_{ae}} \over {\partial\phi^{\alpha}}}\delta^{(3)}(\boldsymbol{x}-\boldsymbol{y})
\end{eqnarray}

\noindent
The integrated form of the mixed partials condition enables is given by \cite{EYO}

\begin{eqnarray}
\label{PARTIALL}
\pi_{\alpha}=f_{\alpha}(\phi^{\beta})-{i \over G}
\int_{\Gamma}\delta{A}^a_iB^i_e{{\partial\Psi_{ae}} \over {\partial\phi^{\alpha}}}
\end{eqnarray}

\noindent
where $f_{\alpha}$ is a function entirely of the matter fields $\phi^{\alpha}$, which forms a boundary condition on the semiclassical matter momentum $\pi_{\alpha}$ when gravity is turned 
off.\footnote{This function can in principle be chosen to correspond to the observable limit below the Planck scale corresponding to a finite state of quantum gravity.}  
Equation (\ref{PARTIALS2}),(\ref{PARTIALL}) can simplify the computation and interpretation of some of the vertices in the loop expansion of the norm of $\Psi_{GKod}$.\par
\indent  
Equation (\ref{ANZ3}) forms the basic building blocks for construction of the entire series expansion for the generalized Kodama state $\Psi_{GKod}$ in that given a reference configuration for $\Psi_{GKod}$, one can calculate all 1PI vertices with respect to this configuration by repeatedly taking functional derivatives, and then expand 
$\Psi_{GKod}$ for an arbitrary configuration about it provided that the CDJ matrix elements $\Psi_{ae}$ are known.  The set of all 1PI vertices also forms the building blocks for the partition function of the generalized Kodama states $\Psi_{GKod}$, as for any theory.  Therefore, the CDJ Ansatz combined with the mixed partials 
condition (\ref{PARTIALL}) constitute the fundamental building blocks for the computation of this partition function, and consequently the norm of the generalized Kodama states, due to the semiclassical-quantum correspondence.\par
\indent
Before moving on to compute the vertices let us put into context the terms `finite' and `renormalizable' with regard to the partition function.  Finiteness can refers to the state 
$\Psi_{GKod}$ itself (\ref{ANZ}) devoid of any field-theoretical infinities as presented in \cite{EYO}.  Renormalizability at the level of the state is argued by equivalence of the canonical to the path integration approach to quantization \cite{WAVE3}, since the CDJ matrix $\Psi_{ab}$ determining $\Psi_{GKod}$ can be expanded in a `dimensionless' constant $G\Lambda$ \cite{EYOFULL}.\footnote{Though there is no factor of Planck's constant $\hbar$ involved in this expansion.}  Renormalizability and finiteness have an independent meaning for the norm of $\Psi_{GKod}$ when one computes the partition function.  As we will show, this involves a loop expansion in the `dimensionless' coupling constant 
$\hbar{G}\Lambda$ which is the same as that for the pure Kodama state \cite{CSPT}.  Since Chern--Simons theory is renormalizable and finite, then one might be able to infer the same for the partition function for $\Psi_{GKod}$ provided that $\Psi_{GKod}$ exists and is finite.  The pure Kodama state can be written as

\begin{eqnarray}
\label{PARTIALS4}
\Psi_{Kod}[A]=e^{-6(\hbar{G}\Lambda)^{-1}I_{CS}[A(\Sigma)]}.
\end{eqnarray}

\noindent
An Ansatz for the corresponding norm might then be given by

\begin{eqnarray}
\label{PARTIALS4}
\bigl\vert\Psi_{Kod}\bigr\vert^{2}=
\bigl<\Psi_{Kod}\bigl\vert\Psi_{Kod}\bigr>
=\int{DA}\bigl<\Psi_{Kod}\bigl\vert{A}\bigr>\bigl<{A}\bigr\vert\Psi_{Kod}\bigr>\nonumber\\
=\prod_{x,a,\alpha}\int{dA}^a_i(x)\Psi^{*}_{Kod}[A]\Psi_{Kod}[A].
\end{eqnarray}

\noindent
Equation (\ref{PARTIALS4}) forms the basis for Chern-Simons perturbation theory with dimensionless coupling constant $\sqrt{\hbar{G}\Lambda}$.  Since the Chern--Simons functional plays the role of the wavefunction of the universe for pure four-dimensional gravity with $\Lambda$ term, 
then (\ref{PARTIALS4}) might be used to define the norm.  It is apriori known, in light of the complex nature of the Ashtekar variables, that an appropriate contour in the complex plane in connection space must be chosen in order to guarantee convergence of
(\ref{PARTIALS4}) as shown in \cite{CSPT}.\par  
\indent
We would like to extend the same procedure to the generalized Kodama states (\ref{ANZ}), which would imply the following definition, subject to the appropriate contour of integration over the gravitational variables

\begin{eqnarray}
\label{PARTIALS3}
\bigl\vert\Psi_{GKod}\bigr\vert^{2}=
\bigl<\Psi_{GKod}\bigl\vert\Psi_{GKod}\bigr>
=\int{DA}{D\phi}\bigl<\Psi_{Kod}\bigl\vert{A},\phi\bigr>\bigl<{A},\phi\bigr\vert\Psi_{Kod}\bigr>\nonumber\\
\bigl\vert\Psi_{GKod}\bigr\vert^{2}=\prod_{x,i,a,\alpha}\int{dA}^a_i(\boldsymbol{x})
d\phi^{\alpha}(\boldsymbol{x})\Psi^{*}_{GKod}[A,\phi]\Psi_{GKod}[A,\phi],
\end{eqnarray}

\noindent
for the corresponding norm.  In (\ref{PARTIALS3}) the requirement to assess the convergence properties as well as the renormalizability of the configuration space integral presents itself.  In a sense the 
issues of finiteness and a true test of the renormalizability of our quantization procedure have been shifted from the level of the state, for which the canonical approach already implies convergence of the path integral by the arguments of \cite{WAVE3}, to the level of its norm, for which such issues must now be re-examined.  If the coupling constant for (\ref{PARTIALS3}) is the same as that 
for (\ref{PARTIALS4}), then one might be able to infer renormalizability of (\ref{PARTIALS3}) as a loop expansion in this coupling constant.  Lastly, the expectation value of observables for the pure an generalized Kodama states can be postulated in terms of the partition function, by

\begin{eqnarray}
\label{PAP5}
\bigl<O\bigr>
={{\int{DA}\Psi^{*}_{Kod}[A]\hat{O}[A]\Psi_{Kod}[A]} \over {\int{DA}\Psi^{*}_{Kod}[A]\Psi_{Kod}[A]}};~~
\bigl<O\bigr>={{\int{DA}\Psi^{*}_{GKod}[A]\hat{O}[A]\Psi_{GKod}[A]} \over {\int{DA}\Psi^{*}_{GKod}[A]\Psi_{GKod}[A]}}
\end{eqnarray}

\section{Computation of the 1PI vertices for $\Psi_{GKod}$.}

In order to have a chance of verifying the renormalizability renormalizability of the generalized Kodama partition function, it might sufficient to verify that the coupling constant 
for $\Psi_{GKod}$ is indeed dimensionless as it is for $\Psi_{Kod}$.  We will demonstrate equality of the coupling constants by expansion of some of the 1PI vertices for the generalized Kodama states about their pure Kodama counterparts.  Additionally, we will provide some interesting geometric interpretations of these vertices by analogy to the Kaluza--Klein theory of infinite dimensional functional spaces.\footnote{For lack of a more descriptive term.}\par
\indent  
We will now compute some of the vertices that will appear in the loop expansion of the partition function for the generalized Kodama state $\Psi_{GKod}$.\footnote{We emphasize that the 1PI vertices are uniquely determined to all orders once the CDJ matrix $\Psi_{ab}$ and the semiclassical matter momentum $\pi_{\alpha}$ are known.  These quantities can be determined only
by solving the canonical constraints in conjunction with inputting the proper semiclassical limit.  Hence once solved, the existence of these elements $\Psi_{ab}$ can be considered equivalent to the existence of the norm of $\Psi_{GKod}$, as well as to $\Psi_{GKod}$ itself.}  Consider the expansion about a reference configuration $(\alpha,\varphi)$ of the effective action 
defining $\Psi_{GKod}$.  In shorthand notation,

\begin{eqnarray}
\label{PARTI5}
\Gamma[\alpha+A,\varphi+\phi]\nonumber\\
=\Gamma_{0}[\alpha,\varphi]+\Gamma^i_{a}A^a_{i}+\Gamma_{\alpha}\phi^{\alpha}
+{1 \over {2!}}\bigl(\Gamma^{ij}_{ab}A^a_{i}A^j_{b}+2(\Gamma_{\alpha})^i_{a}A^a_{i}\phi^{\alpha}
+\Gamma_{\alpha\beta}\phi^{\alpha}\phi^{\beta}\bigr)\nonumber\\
+{1 \over {3!}}\Bigl(\Gamma^{ijk}_{abc}A^a_{i}A^b_{j}A^c_{k}
+3(\Gamma_{\alpha})^{ij}_{ab}A^a_{i}A^b_{j}\phi^{\alpha}
+3(\Gamma_{\alpha\beta})^i_{a}A^a_{i}\phi^{\alpha}\phi^{\beta}
+\Gamma_{\alpha\beta\gamma}\phi^{\alpha}\phi^{\beta}\phi^{\gamma}\Bigr)+\dots
\end{eqnarray}

\noindent
Equation (\ref{PARTI5}) can be separated into three contributions: A pure gravitational contribution, a pure matter contribution and a mixed contribution.  The zeroth-order vertex $\Gamma_0
=\Gamma^{(0)}_{\Sigma}[\alpha,\varphi]$ corresponds to the action corresponding to the full solution of the quantum constraints of gravity coupled to matter, projected onto the final 3-surface and evaluated for some arbitrarily chosen reference configuration $(\alpha,\varphi)$ of the gravitational and matter fields.  Let us first compute the lowest order vertices, which we will need in 
order to evaluate the Gaussian part of the norm for $\Psi_{GKod}$.\par
\indent  
The pure gravitational contribution is given by

\begin{eqnarray}
\label{PARTI6}
I_{grav}(A)=\Gamma^i_{a}A^a_{i}+{1 \over {2!}}\Gamma^{ij}_{ab}A^a_{i}A^j_{b}
+{1 \over {3!}}\Gamma^{ijk}_{abc}A^a_{i}A^b_{j}A^c_{k}+\dots,
\end{eqnarray}

\noindent
The pure matter conribution is given by

\begin{eqnarray}
\label{PARTI7}
I_{matter}=\Gamma_{\alpha}\phi^{\alpha}+{1 \over {2!}}\Gamma_{\alpha\beta}\phi^{\alpha}\phi^{\beta}
+{1 \over {3!}}\Gamma_{\alpha\beta\gamma}\phi^{\alpha}\phi^{\beta}\phi^{\gamma}+\dots
\end{eqnarray}

\noindent
and the mixed terms, are given by

\begin{eqnarray}
\label{PARTI8}
I_{mixed}=(\Gamma_{\alpha})^i_{a}A^a_{i}\phi^{\alpha}
+{1 \over 2}(\Gamma_{\alpha})^{ij}_{ab}A^a_{i}A^b_{j}\phi^{\alpha}
+{1 \over 2}(\Gamma_{\alpha\beta})^i_{a}A^a_{i}\phi^{\alpha}\phi^{\beta}+\dots
\end{eqnarray}

\noindent
Let us now substitute the CDJ Ansatz into (\ref{PARTI6}), (\ref{PARTI7}) and (\ref{PARTI8}).\par
\noindent

\subsection{Pure gravitational vertices}

I) Starting with $I_{grav}$,

\begin{eqnarray}
\label{PARTI9}
I_{grav}=(\Psi_{ae}B^i_{e})\Bigl\vert_{\alpha}A^a_{i}
+{1 \over {2!}}\Bigl({\delta \over {\delta\alpha^b_j}}(\Psi_{ae}B^i_{e})\Bigr)A^a_{i}A^b_{j}
+{1 \over {3!}}\Bigl({\delta \over {\delta\alpha^c_k}}{\delta \over {\delta\alpha^b_j}}(\Psi_{ae}B^i_{e})\Bigr)A^a_{i}A^b_{j}A^c_{k}+\dots
\end{eqnarray}

\noindent
Substituting the Ansatz $\Psi_{ae}=-(6/\Lambda)\bigl(\delta_{ae}
+(\Lambda/6)\epsilon_{ae}\bigr)$ which expands $\Psi_{GKod}$ about $\Psi_{Kod}$ using the CDJ deviation matrix $\epsilon_{ae}$ as in \cite{EYOFULL}, we have

\begin{eqnarray}
\label{PARTI10}
I_{grav}=-{6 \over \Lambda}\biggl(B^i_{a}[\alpha]A^a_i+{1 \over {2!}}D^{ij}_{ab}(\boldsymbol{x},\boldsymbol{y})A^a_i(x)A^b_j(y)+{1 \over {3!}}\epsilon^{ijk}_{abc}A^a_iA^b_jA^c_k\nonumber\\
+{\Lambda \over 6}\Bigl[A^a_{i}+{1 \over {2!}}A^a_{i}A^b_{j}{\delta \over {\delta\alpha^b_j}}
+{1 \over {3!}}A^a_{i}A^b_{j}A^c_{k}{\delta \over {\delta\alpha^c_k}}{\delta \over {\delta\alpha^b_j}}+\dots\Bigr](\epsilon_{ea}B^i_e[\alpha])\Bigr)~~~~\nonumber\\
\end{eqnarray}

\noindent
A check of the mass dimensions

\begin{eqnarray}
B^i_{a}(\alpha)A^a_i=\int_{\Sigma}d^3{x}~B^i_{a}(\alpha(x))A^a_i(x)
\end{eqnarray}

\noindent
yields $[\int{d^3x}]+[B]+[A]=-3+2+1=0$, and likewise for the higher order terms, which means that $I_{grav}$ is of mass dimension $[I_{grav}]=-2$.\par
\indent
Let us now acquire some intuition into the geometric nature of these vertices.  Starting from the 1-point vertex for the connection,

\begin{equation}
\label{VER1}
\Gamma^i_{a}(x)={{\delta\Gamma_{\Sigma}} \over {\delta{A}^a_i(x)}}=\Psi_{ae}(x)B^i_{e}(x)
\end{equation}

\noindent
(\ref{VER1}) is just the CDJ Ansatz.  One may choose to expand about a configuration for which this term is zero, but this restriction is not compulsory.  To see more clearly the relation to the pure Kodama state and its renormalizability, let us express (\ref{VER1}) in terms of the CDJ deviation matrix 
$\epsilon_{ae}$.  This gives

\begin{eqnarray}
\label{DEVI}
\Gamma^i_a(x)=-{6 \over \Lambda}B^i_e(x)\bigl(\delta_{ae}+{\Lambda \over 6}\epsilon_{ae}(x)\bigr).
\end{eqnarray}

\noindent
Equation(\ref{VER1}) can be seen as a correction to the Chern-Simons one-point vertex (or curvature), due to gravity-matter quantum effects encoded in $\epsilon_{ae}$.\footnote{Note by \cite{EYORENORM} that $\epsilon_{ae}\sim\epsilon_{ae}[G\Lambda]$ can be expanded in an power series in the `Dimensionless' coupling constant $G\Lambda$ with no factor of $\hbar$.}  Moving on to the 2-point connection vertex, we have

\begin{eqnarray}
\label{VER11}
\Gamma^{ij}_{ab}(x,y)={{\delta^{2}\Gamma_{\Sigma}} \over {\delta{A}^a_i(x){\delta{A}^b_j(y)}}}
\biggl\vert_{A=\alpha,\phi=\varphi}
={{\delta\Psi_{ae}(x)} \over {\delta{A}^b_j(y)}}B^i_{e}(x)+\Psi_{ae}(x){{\delta{B^i_e}(x)} \over {\delta{A}^b_j(y)}}\nonumber\\
=\Bigl(\Psi_{ae}D^{ij}_{eb}+{{\partial\Psi_{ae}} \over {\partial{A}^b_j}}B^i_{e}\Bigr)\delta^{(3)}(\boldsymbol{x}-\boldsymbol{y})\biggl\vert_{A=\alpha,\phi=\varphi}
\end{eqnarray}

\noindent
which can be seen as a modification of the Chern--Simons inverse propagator due to gravity-matter effects.  In this respect, (\ref{VER11}) can be written as 

\begin{eqnarray}
\label{DEVI1}
\Gamma^{ij}_{ab}(x,y)=\Psi_{ae}\bigl(D^{ij}_{eb}(\boldsymbol{x},\boldsymbol{y})+(\Psi^{-1})_{eg}{{\partial\Psi_{gf}} \over {\partial{A}^b_j}}B^i_{f}\bigr)\delta^{(3)}(\boldsymbol{x}-\boldsymbol{y})
\end{eqnarray}

\noindent
(\ref{VER11}) and (\ref{DEVI1}) represent a `generalized' kinetic operator or inverse propagator, twisted due to the presence of the matter fields.  The following notation 
pertains for the connection `propagator'

\begin{equation}
\label{PROPA}
\Delta^{ab}_{ij}(\boldsymbol{x},\boldsymbol{y})=\bigl(\Gamma^{ij}_{ab}(\boldsymbol{x},\boldsymbol{y})\bigr)^{-1}
\end{equation}

\noindent
such that

\begin{equation}
\label{PROPA1}
\int_{\Sigma}d^3{z}\Delta^{ab}_{ij}(\boldsymbol{x},\boldsymbol{z})\Gamma^{jk}_{bc}(\boldsymbol{z},\boldsymbol{y})=\delta^{k}_{i}\delta^a_{c}\delta^{(3)}(\boldsymbol{x}-\boldsymbol{y})
\end{equation}

\noindent
Note that for $\Psi_{ae}=-(6/\Lambda)\delta_{ae}$, (\ref{VER11}) would correspond exactly to the kinetic operator for $\Psi_{Kod}$, which is the same as for Chern--Simons partition function.  This is related to the linking number between two loops in the loop representation, as in

\begin{equation}
\label{PROPA2}
\oint_{\gamma}{dx^i}\oint_{\gamma}{dy^j}\Delta^{ab}_{ij}(x,y)\tau_{a}\tau_{b}\biggl\vert_{\Psi_{ab}=\delta_{ab}}
=\oint_{\gamma}{dx^i}\oint_{\gamma}{dy^j}\epsilon_{ijk}{{x^k-y^k} \over {\vert{\boldsymbol{x}-\boldsymbol{y}}\vert^3}}
\end{equation}

\noindent
In its relation to the pure Kodama state, (\ref{VER11}) can also be written 

\begin{eqnarray}
\label{DEVI2}
\Gamma^{ij}_{ab}(x,y)=-{6 \over \Lambda}D^{kj}_{fb}\Bigl[\delta^i_{k}\delta_{af}
+{\Lambda \over 6}\Bigl(\delta^i_{k}\delta^e_{f}+(D^{-1})^{gf}_{lk}B^i_{e}{\partial \over {\partial{A}^g_l}}\Bigr)\epsilon_{ae}\Bigr]
\delta^{(3)}(\boldsymbol{x}-\boldsymbol{y})
\end{eqnarray}

\noindent
Another way to represent the gravitational propagator, seen as an instantaneous Coulombic interaction, is

\begin{eqnarray}
\label{GRAVPROP}
\Delta^{ab}_{ij}(\boldsymbol{x},\boldsymbol{y})
=\bigl<{0}_F\bigl\vert\hat{A}^a_i(\boldsymbol{x})\hat{A}^b_j(\boldsymbol{y})\bigl\vert{0}_F\bigr>.
\end{eqnarray}

\noindent
Equation (\ref{GRAVPROP}) is the amplitude for a particle associated with the connection to propagate from spacetime position $x=(\boldsymbol{x},t)$ to another position at the same time 
$y=(\boldsymbol{y},t)$.\footnote{Hence there is a nonzero probability for fields to propagate outside the lightcone with respect to some vacuum state $\bigl\vert{0}_F\bigr>$.  Incidentally, the vaccum state corresponds to the reference field configuration $(\alpha,\varphi)$ chosen.}  Another convenient interpretation for the kinetic operator, in line with the concepts of \cite{GEOMETRY}, is the covariant metric on the space of gravitational field configurations.  Conversely, the propagator would be the analogue to the contravariant metric.\par
\indent
The 3-point vertex is given by

\begin{eqnarray}
\label{VER1111}
\Gamma^{ijk}_{abc}(x,y,z)={{\delta^{3}\Gamma_{\Sigma}} \over {\delta{A}^a_i(x){\delta{A}^b_j(y)}{\delta{A}^c_k(x)}}}\biggl\vert_{A=\alpha,\phi=\varphi}\nonumber\\
=\Bigl[\Psi_{ae}\epsilon^{ijk}_{ebc}+{{\partial\Psi_{ae}} \over {\partial{A}^c_{k}}}D^{ij}_{eb}
+{{\partial\Psi_{ae}} \over {\partial{A}^b_{j}}}D^{ik}_{ec}
+B^i_{e}{{\partial^{2}\Psi_{ae}} \over {\partial{A}^b_{j}
\partial{A}^c_{k}}}\Bigr]\delta^{(3)}(\boldsymbol{z}-\boldsymbol{x})\delta^{(3)}(\boldsymbol{x}-\boldsymbol{y})
\biggl\vert_{A=\alpha,\phi=\varphi}
\end{eqnarray}

\indent
The first term on the last line of (\ref{VER1111}), for an isotropic CDJ matrix $\Psi_{ae}$, corresponds to the trivalent vertex in Chern--Simons perturbation theory.  To expose the relationship to the pure Kodama state, again one can expand this in terms of the CDJ deviation matrix

\begin{eqnarray}
\label{VER111}
\Gamma^{ijk}_{abc}(x,y,z)
=-{6 \over \Lambda}\Bigl[\epsilon^{ijk}_{abc}
+{\Lambda \over 6}\Bigl(\epsilon^{ijk}_{ebc}+D^{ij}_{eb}{\partial \over {\partial{A}^c_{k}}}\nonumber\\
+D^{ik}_{ec}{\partial \over {\partial{A}^b_{j}}}
+B^i_{e}{{\partial^{2}} \over {\partial{A}^b_{j}\partial{A}^c_{k}}}\Bigr)
\epsilon_{ae}\Bigr]
\delta^{(3)}(\boldsymbol{z}-\boldsymbol{x})\delta^{(3)}(\boldsymbol{x}-\boldsymbol{y})
\end{eqnarray}

\par
\indent
Simplifying (\ref{VER111}) further, we obtain

\begin{eqnarray}
\label{VER112}
\Gamma^{ijk}_{abc}(x,y,z)
=\delta^{(3)}(\boldsymbol{z}-\boldsymbol{x})\delta^{(3)}(\boldsymbol{x}-\boldsymbol{y})
\biggl[-{6 \over \Lambda}\epsilon^{ijk}\epsilon_{ebc}\bigl(\delta_{ae}+{\Lambda \over 6}\epsilon_{ae}\bigr)\nonumber\\
-\Bigl[\epsilon^{ijl}\epsilon_{ebf}D^{ij}_{eb}{\partial \over {\partial{A}^c_k}}
+\epsilon^{ikl}\epsilon_{ecf}D^{ik}_{ec}{\partial \over {\partial{A}^b_j}}
+B^i_{e}{{\partial^{2}} \over {\partial{A}^b_{j}\partial{A}^c_{k}}}\Bigr]\epsilon_{ae}\biggr]\biggl\vert_{A=\alpha,\phi=\varphi}
\end{eqnarray}

\noindent
from which it is clear that the first term is the gravitational Chern--Simons trivalent vertex corrected by matter effects of order $G\Lambda$, followed by additional corrections.  The process can be continued to any order desired.  The effect of coupling matter to gravity is to create an infinite number of higher-order vertices.  From this perspective it is more
complicated than the Chern--Simons functional, which is restricted to just the trivalent vertex.  On the other hand, all higher-order vertices are uniquely determined from the CDJ Ansatz.  From this perspective, all the information of these vertices is encoded in the first-order vertex.\footnote{Hence, the existence of the expansion to all orders is guaranteed due to the existence of $\Psi_{GKod}$ as shown in \cite{EYO}.  This is in stark contrast to nonrenormalizable theories, which require an infinite number of parameters to determine the quantum effective action from its classical version.}\par
\indent

\subsection{Pure matter vertices}

\noindent
Moving on to the pure matter vertices,

\begin{eqnarray}
\label{PARTI11}
I_{matter}=
iG\Bigl(\pi_{\alpha}\Bigl\vert_{\varphi}\phi^{\alpha}
+{1 \over {2!}}\Bigl({{\delta\pi_{\alpha}} \over {\delta\varphi^{\beta}}}\Bigr)\phi^{\alpha}\phi^{\beta}
+{1 \over {3!}}\Bigl({{\delta\pi_{\alpha}} \over {\delta\varphi^{\beta}{\delta\varphi^{\gamma}}}}\Bigr)\phi^{\alpha}\phi^{\beta}\phi^{\gamma}+...\Bigr)\nonumber\\
={6 \over \Lambda}\Bigl[{{iG\Lambda} \over 6}\Bigl(\pi_{\alpha}\Bigl\vert_{\varphi}\phi^{\alpha}
+{1 \over {2!}}\biggl({{\delta\pi_{\alpha}} \over {\delta\varphi^{\beta}}}\Bigr)\phi^{\alpha}\phi^{\beta}
+{1 \over {3!}}\Bigl({{\delta\pi_{\alpha}} \over {\delta\varphi^{\beta}{\delta\varphi^{\gamma}}}}\Bigr)\phi^{\alpha}\phi^{\beta}\phi^{\gamma}+...\biggr).\nonumber\\
\end{eqnarray}

\noindent
We have written (\ref{PARTI11}) in the form indicated to combine it with $I_{grav}$.  Note that the coefficient of $6/\Lambda$ is dimensionless, since $G\Lambda$ is dimensionless as is every term in brackets.  For example, the first term is given by

\begin{eqnarray}
\label{PARTI12}
\pi_{\alpha}\Bigl\vert_{\varphi}\phi^{\alpha}=\int_{M}d^3{x}~\pi_{\alpha}[\varphi(x)]
\phi^{\alpha}(x)
\end{eqnarray}

\noindent
giving $[\int{d^3}x]+[\pi]+[\phi]=-3+2+1=0$, and likewise for the remaining terms, consistent with mass dimensions $[I_{matter}]=-2$.\par
\noindent
The first-order (1 point) pure matter vertex is given by\footnote{We will for notational convenience take all derivatives as evaluated on the reference field configuration $A=\alpha$, $\phi=\varphi$ unless specified otherwise.}

\begin{equation}
\label{VER2}
\Gamma_{\alpha}(x)={{\delta\Gamma_{\Sigma}} \over {\delta\phi^{\alpha}(x)}}=iG\pi_{\alpha}(x)\biggl\vert_{A=\alpha,\phi=\varphi}
\end{equation}

\noindent
Equation (\ref{VER2}) is the matter analogue to the CDJ Ansatz, which also satisfies the SQC.  The right hand side is the Ansatz for the action of the matter momentum operator on $\Psi_{GKod}$, the analogue of (\ref{NORM1})

\begin{equation}
\label{NORM2}
-i\hbar{\delta \over {\delta\phi^{\alpha}(x)}}\Psi_{GKod}=\pi(x)\Psi_{GKod}
\end{equation}

\noindent
Moving on to the 2-point matter vertex, we have

\begin{equation}
\label{VER22}
\Gamma_{\alpha\beta}(x,y)={{\delta^2\Gamma_{\Sigma}} \over {\delta\phi^{\alpha}(x){\delta\phi^{\beta}(y)}}}
=iG{{\delta\pi_{\alpha}(x)} \over {\delta\phi^{\beta}(y)}}=iG{{\partial\pi_{\alpha}} \over {\partial\phi^{\beta}}}\delta^{(3)}(\boldsymbol{x}-\boldsymbol{y})
\end{equation}

\noindent
(\ref{VER22}) has the interpretation of the matter-field kinetic term or inverse propagator.  In analogy to the connection propagator,

\begin{equation}
\label{PROPA3}
\Delta^{\alpha\beta}(x,y)=\bigl(\Gamma_{\alpha\beta}(x,y)\bigr)^{-1}
\end{equation}

\noindent
such that

\begin{equation}
\label{PROPA4}
\int_{\Sigma}d^{3}z~\Delta^{\alpha\beta}(x,z)\Gamma_{\beta\gamma}(z,y)=\delta^{\alpha}_{\gamma}\delta^{(3)}(\boldsymbol{x}-\boldsymbol{y})
\end{equation}

\noindent
Again, (\ref{PROPA3}) can be seen as contravariant and covariant metrics on the space of matter field configurations.  In analogy to the gravitational case, one can also write the matter propagator in terms of expectation values

\begin{eqnarray}
\label{MATTERPROP}
\Delta^{\alpha\beta}(\boldsymbol{x},\boldsymbol{y})
=\bigl<0_F\bigl\vert\hat{\phi}^{\alpha}(\boldsymbol{x})\hat{\phi}^{\beta}(\boldsymbol{y})
\bigr\vert{0}_F\bigr>
\end{eqnarray}

\noindent
which as well signifies the propagation with respect to the reference vacuum configuration between two spatial points at the same time $t$ via an instantaneous Coulombic-type interation.  The matter kinetic operator also has another interesting interpretation.  Taking the trace of (\ref{VER22}) and sets $\boldsymbol{x}=\boldsymbol{y}$, we get

\begin{equation}
\label{INTERP}
\hbox{tr}\Gamma_{\alpha\beta}(x,y)=\sum_{\alpha}\Gamma^{\alpha}_{\alpha}(x,x)=iG{{\partial\pi_{\alpha}} \over {\partial\phi^{\alpha}}}\delta^{(3)}(0)=iG\delta^{(3)}(0)q.
\end{equation}

\noindent
Equation (\ref{INTERP}) is the divergence of the matter momentum at the point $x$.  This is none other than $\Omega_1$, the matter contribution to the first order of singularity of the Hamiltonian constraint, which acts as a source for the functional divergence of the CDJ matrix 
elements \cite{EYOFULL}.  This is a quantum effect, also necessary to compute the norm of 
$\Psi_{GKod}$, which manifests itself at the linearized level of the constraints.\par
\indent

\subsection{Mixed vertices}

Moving on to the mixed terms, we see that there is a single occurrence of the matter semiclassical conjugate momentum $\pi_{\alpha}$ in each term

\begin{eqnarray}
\label{PARTI13}
I_{cross}=iG\biggl(\Bigl({{\delta\pi_{\alpha}} \over {\delta\alpha^a_i}}\Bigr)A^a_{i}\phi^{\alpha}
+{1 \over 2}\Bigl({{\delta^{2}\pi_{\alpha}} \over {\delta\alpha^a_{i}{\delta\alpha^b_{j}}}}
\Bigr)A^a_{i}A^b_{j}\phi^{\alpha}
+{1 \over 2}\Bigl({{\delta^{2}\pi_{\alpha}} \over {\delta\varphi^{\beta}\delta\alpha^a_i}}\Bigr)A^a_{i}\phi^{\alpha}\phi^{\beta}+...\biggr).
\end{eqnarray}

\noindent
For the purposes of norm computation we can choose either to combine the cross terms with the matter terms (\ref{PARTI11}), which (\ref{PARTI13}) is already suited to do, or to combine them with the gravitational terms (\ref{PARTI10}).  The latter requires use of the mixed partials condition and leads to

\begin{eqnarray}
\label{PARTI14}
I_{cross}=
\Bigl(B^i_{e}{{\delta\Psi_{ae}} \over {\delta\phi^{\alpha}}}\Bigr)A^a_{i}\phi^{\alpha}
+{1 \over 2}\Bigl({{\delta^{2}(\Psi_{be}B^j_e)} \over {{\delta\alpha^a_i}{\delta\phi^{\alpha}}}}\Bigr)A^a_{i}A^b_{j}\phi^{\alpha}
+{1 \over 2}\Bigl({{\delta^{2}(B^i_{e}\Psi_{ae})} \over {{\delta\varphi^{\beta}\delta{\varphi^{\alpha}}}}}\Bigr)A^a_{i}\phi^{\alpha}\phi^{\beta}+...
\end{eqnarray}

\noindent
As a check on the mass dimensions, an examination of the first term of (\ref{PARTI14}) yields

\begin{eqnarray}
\Bigl(B^i_{e}{{\delta\Psi_{ae}} \over {\delta\phi^{\alpha}}}\Bigr)A^a_{i}\phi^{\alpha}
=\int_{\Sigma}d^3{x}\int_{\Sigma}d^3{y}\Bigl(B^i_{e}(x){{\delta\Psi_{ae}(x)} \over {\delta\varphi^{\alpha}(y)}}\Bigr)A^a_{i}(x)\phi^{\alpha}(y).
\end{eqnarray}

\noindent
We must not forget the mass dimension of $-1$ due to the $\varphi$ in the denominator.  Thus the total is
$[\int\int{d}^6{x}]+[B]+[\Psi]-[\varphi]+[A]+[\phi]+[\delta^{(3)}(x-y)]=-6+2-2-1+1+1+3=-2$, which makes $[I_{cross}]=-2$.  So all quantities have been placed on the same footing.  Our 
ultimate goal is to divide $I=I_{grav}+I_{matter}+I_{cross}$ by $\hbar{G}$ in order to extract the dimensionless constant $\hbar{G}\Lambda$.  This would mean that the partition function for the generalized Kodama state $\Psi_{GKod}$ is renormalizable.  But first, we must express 
(\ref{PARTI14}) in terms of the CDJ deviation matrix via the Ansatz 
$\Psi_{ae}=-(6/\Lambda)\bigl(\delta_{ae}+(\Lambda/6)\epsilon_{ae}\bigr)$.  Let us compute three terms of the series to illustrate the idea.\par
\noindent
(IIIa) Starting with the first term of (\ref{PARTI14})

\begin{eqnarray}
\label{PARTI15}
\Bigl(B^i_{e}{{\delta\Psi_{ae}} \over {\delta\phi^{\alpha}}}\Bigr)A^a_{i}\phi^{\alpha}
=-{6 \over \Lambda}\Bigl[{\Lambda \over 6}
\Bigl(B^i_{e}{{\delta\Psi_{ae}} \over {\delta\phi^{\alpha}}}\Bigr)A^a_{i}\phi^{\alpha}\Bigr]
\end{eqnarray}

\noindent
(IIIb) Moving on to the second term of (\ref{PARTI14}),

\begin{eqnarray}
\label{PARTI16}
{1 \over 2}\Bigl({{\delta^{2}(\Psi_{be}B^j_e)} \over {{\delta\alpha^a_i}{\delta\phi^{\alpha}}}}\Bigr)A^a_{i}A^b_{j}\phi^{\alpha}
=-{6 \over \Lambda}\Bigl[{\Lambda \over {12}}
\Bigl({\delta \over {\delta\alpha^a_i}}
{\delta \over {\delta\phi^{\alpha}}}+D^{ji}_{ea}(\alpha){\delta \over {\delta\varphi^{\alpha}}}\Bigr)\epsilon_{be}B^j_{e}\Bigr]A^a_{i}A^b_{j}\phi^{\alpha}
\end{eqnarray}

\noindent
where we have used the independence of $\delta_{ae}$ as well as the gravitational variable $B^i_a$ upon $\phi^{\alpha}$.  (IIIc) Moving on to the third term of (\ref{PARTI14}),

\begin{eqnarray}
{1 \over 2}\Bigl({{\delta^{2}(B^i_{e}\Psi_{ae})} \over {{\delta\varphi^{\beta}\delta{\varphi^{\alpha}}}}}\Bigr)A^a_{i}\phi^{\alpha}\phi^{\beta}
=-{6 \over \Lambda}\Bigl[{\Lambda \over {12}}
B^i_{e}\Bigl({{\delta^{2}\epsilon_{ae}} \over {\delta\varphi^{\beta}\delta{\varphi^{\alpha}}}}\Bigr)\Bigr]A^a_{i}\phi^{\alpha}\phi^{\beta}
\end{eqnarray}

\noindent
and so on and so forth.\par
\indent
Combining all terms, we find that

\begin{eqnarray}
\label{DIMLESS}
\Gamma_{\Sigma}=I=I_{grav}+I_{cross}+I_{matter}=-{6 \over \Lambda}I_{\Sigma}
\end{eqnarray}

\noindent
where $I_{\Sigma}$ is dimensionless.  Since the generalized Kodama state is given by

\begin{eqnarray}
\label{DIMLESS1}
\Psi_{GKod}=e^{(\hbar{G})^{-1}\Gamma_{\Sigma}}=e^{-6(\hbar{G}\Lambda)^{-1}I_{\Sigma}},
\end{eqnarray}

\noindent
we find that the coupling constant for the partition function for $\Psi_{GKod}$, 
$k=\sqrt{(\hbar{G}\Lambda/6)}$ is not only dimensionless, but also very small.  We have
anticipated rescaling the quadratic term in $A$ and $\phi$ in order to transfer this coupling constant to the cubic and all higher order self-interaction vertices.\par
\indent
Let us now compute some mixed vertices to acquire a geometric interpretation.  Acting on (\ref{VER1}),  

\begin{equation}
\label{VER12}
{{\delta\Gamma^i_{a}(x)} \over {\delta\phi^{\alpha}(y)}}\equiv(\Gamma_{\alpha}(x,y))^i_a= {{\delta^{2}\Gamma_{\Sigma}} \over {\delta{A}^a_i(x){\delta\phi^{\alpha}(y)}}}
=B^i_{e}{{\partial\Psi_{ae}} \over {\partial\phi^{\alpha}}}\delta^{(3)}(\boldsymbol{x}-\boldsymbol{y})
\end{equation}

\noindent
Acting on (\ref{VER2})

\begin{equation}
\label{VER21}
{{\delta\Gamma_{\alpha}(x)} \over {\delta{A}^a_i(y)}}={{\delta^{2}\Gamma_{\Sigma}} \over {\delta{A}^a_i(y)\delta\phi^{\alpha}(x)}}
=(iG){{\partial\pi_{\alpha}} \over {\partial{A}^a_i}}\delta^{(3)}(\boldsymbol{x}-\boldsymbol{y})
\end{equation}

\noindent
Compared with (\ref{PARTIALS2}) and (\ref{VER12}) we see that this is none other than the mixed partials condition.\par
\indent
The low order vertices for the matter-gravity system have an interesting physcial interpretation.  When one views the fields as coordinates in an infinite dimensional functional 
manifold $\Gamma\sim(\Gamma_{\alpha},\Gamma_{\varphi})$, then one can define a length squared functional by

\begin{eqnarray}
\delta{s}^2=\int_{\Sigma}\int_{\Sigma}d^3xd^3y\Bigl[
\Gamma_{\alpha\beta}\delta\varphi(\boldsymbol{x})\delta\varphi(\boldsymbol{y})
+\Gamma^{ij}_{ab}(\boldsymbol{x},\boldsymbol{y})
\delta\zeta^a_i(\boldsymbol{x})\delta\zeta^b_j(\boldsymbol{y})\Bigr]
\end{eqnarray}

\noindent
where we have defined the `shifted' functional one-form

\begin{eqnarray}
\label{SHIFT}
\delta\zeta^a_i(\boldsymbol{x})=
\delta{\alpha}^a_i(\boldsymbol{x})+\int_{\Sigma}\int_{\Sigma}d^3yd^3z
\Delta^{ae}_{ik}(\boldsymbol{x},\boldsymbol{y})(\Gamma_{\alpha}(\boldsymbol{y},\boldsymbol{z}))^k_e\delta\phi^{\alpha}(\boldsymbol{z})
\end{eqnarray}

\noindent
It is clear that an isometry in the $(\zeta,\varphi)$ basis is $O(N)\otimes{O}(3)\otimes{O}(3)$.  Likewise, one can define the infinite dimensional vector field corresponding to this one-form,

\begin{eqnarray}
\label{SHIFT1}
{\delta \over {\delta\zeta^{\alpha}(\boldsymbol{x})}}
={\delta \over {\delta\varphi^{\alpha}(\boldsymbol{x})}}
-\int_{\Sigma}\int_{\Sigma}d^3yd^3z
\Delta^{ae}_{ik}(\boldsymbol{x},\boldsymbol{y})(\Gamma_{\alpha}(\boldsymbol{y},\boldsymbol{z}))^k_e
{\delta \over {\delta\alpha^a_i(\boldsymbol{z})}}
\end{eqnarray}

\noindent
This has the interpretation of an infinite dimensional version of Kaluza--Klein theory, in which the higher dimensional space of dimension $9+N$ is split into $9$ dimensions for the gravitational configuration variables $A^a_i$ plus $N$ dimensions for their matter counterparts $\phi^{\alpha}$.  The covariant metric in the $(\delta\alpha,\delta\varphi)$ basis is given by

\begin{displaymath}
g_{IJ}=
\left(\begin{array}{ccc}
\Gamma_{\alpha\beta}-(\Gamma_{\alpha})^a_i\Gamma^{ij}_{ab}(\Gamma_{\beta})^b_j & 
\Delta^{ab}_{ij}(\Gamma_{\beta})^j_b\\
\Delta^{ba}_{ji}(\Gamma_{\alpha})^i_a & \Gamma^{ij}_{ab}\\
\end{array}\right).
\end{displaymath}

\noindent
One can think more fundamentally of the Kaluza--Klein analogy in terms of the 3+1 decomposition of spacetime.  The matter field has the interpretation of the `time' coordinate orthogonal to `spatial' or gravitational hypersurfaces.  The off-diagonal elements correspond to the mixed partials condition, which is analogous to the shift vector in relativity.  This signifies an interaction between gravity and matter.\footnote{From one perspective, the presence of matter modifies the gravitational propagator essentailly `shifting' the mass of the particle.}  Hence, if one views the gravitational and the matter fields as the the coordinates of a particle traveling through functional space, then (\ref{SHIFT}) would correspond to the proper time of the particle.  It would be interesting to carry the analogy further, to compute the connection and the curvature on this infinite dimensional manifold.  Such structures occur in the higher order vertices, which we will not compute in detail here.  All higher-order vertices are given, for the general $n$-point vertex by the general expression

\begin{equation}
\label{VERN}
\bigl(\Gamma_{\alpha_{m+1}...\alpha_{n}}(x_1,...,x_n)\bigr)^{i_1...1_m}_{a_1...a_m}
=\Bigl[\prod_{k=1}^{m+1}\prod_{l={m+1}}^{n}{\delta \over {\delta\phi^{\alpha_k}(\boldsymbol{x}_k)}}
{\delta \over {\delta{A}^{a_l}_{i_l}(\boldsymbol{x}_l)}}\Bigr]\Gamma_{\Sigma}
\end{equation}

\section{Generalized Kodama partition function}

\subsection{Gaussian terms}

\noindent
We are now ready to compute the generalized Kodama partition function.  It may be helpful to refer to the appendix, which shows the computation of the Gaussian part of the norm in a simpler finite dimensional case.  The generalization to field theory is straightforward, and we simply quote the results here.  We would like to compute 

\begin{eqnarray}
\label{PARTT}
\Psi_{eff}[\alpha,\varphi]=\int{D}\mu(A,\phi)\Psi_{GKod}[\alpha,\varphi]
\end{eqnarray}

\noindent
where we have defined $\Psi_{GKod}[\alpha,\varphi]=e^{I_{GKod}[\alpha,\varphi]}$, and

\begin{eqnarray}
I=\Gamma_0+{1 \over 2}A^a_i\Delta^{ij}_{ab}A^b_j+B^i_aA^a_i
+{1 \over 2}\phi^{\alpha}\Delta_{\alpha\beta}\phi^{\beta}+\chi^{\alpha}\phi_{\alpha}
+A^a_i(\Gamma_{\alpha})^i_a\phi^{\alpha}+V[A,\phi]
\end{eqnarray}

\noindent
where $V$ represents the total of all the higher order `interaction' terms.  A convenient physical interpretation of the quadratic part of the partition function is to compute the contravariant metric in the infinite dimensional field space, being mindful of any conditions and restrictions upon the background field configuration necessary for the Gaussian integral to converge.  The full partition function can be written down by inspection from the appendix, and is given by

\begin{eqnarray}
\label{PARTY}
\Psi_{eff}[\alpha,\varphi]\propto\Psi[\alpha,\varphi]
e^{V[\delta/\delta{B},\delta/\delta\chi]}
\hbox{exp}\Bigl[{1 \over 2}B^i_a\Gamma^{ab}_{ij}B^i_a\Bigr]
\hbox{exp}\Bigl[{1 \over 2}\widetilde{\chi}_{\alpha}
\widetilde{\Gamma}^{\alpha\beta}\widetilde{\chi}_{\beta}\Bigr]
\end{eqnarray}

\noindent
where we have defined 

\begin{eqnarray}
\label{PARTY1}
\widetilde{\Gamma}^{\alpha\beta}
=\bigl(\Delta_{\alpha\beta}-(\Gamma_{\alpha})^i_a\Gamma^{ab}_{ij}(\Gamma_{\beta})^j_b\bigr)^{-1};~
\widetilde{\chi}_{\alpha}=\xi_{\alpha}-B^i_a\Gamma^{ae}_{il}(\Gamma_{\alpha})^l_e
\end{eqnarray}

\par
\indent
Let us now write out and interpret each term of (\ref{PARTY}) in the infinite dimensional language of the norm of the generalized Kodama state $\Psi_{GKod}$.\par
\noindent
(i) First zeroth order term: This is the argument of the exponential forming the wavefunction of the universe, as determined by the full solution to the quantized constraints to all orders.  One should in principle be able to calculate this to any order desired in $G\Lambda$ using the techniques in 
\cite{EYOFULL}, assuming that the state is finite.  Since for a given configuration this term should cancel out in the computation of expectation values, its actual value is immaterial.

\begin{equation}
\label{ZER}
C\sim\Gamma_{\Sigma}\Bigl\vert_{A=\alpha;\phi=\varphi}
\end{equation}

\noindent
Note that this reference needn't correspond to a critical point of the three-dimensional 
action $I_{\Sigma}$.\par
\noindent
(ii) Second zeroth order term:  This term is the analogue of the Maxwell ground state, and corresponds to a nonlocal Gaussian function

\begin{equation}
\label{ZER1}
{1 \over 2}(a_1)^{-1}_{ij}(b_1)^{i}(b_1)^{j}
\sim\int_{\Sigma}\int_{\Sigma}{d}^3{x}d^3{y}\Gamma^{i}_{a}(x)(\Gamma^{-1})^{ab}_{ij}(x-y)\Gamma^{j}_{b}(y).
\end{equation}

\noindent
The interpretation of (\ref{ZER1}) is that the equilibrium value of the magnetic field about which quantum fluctuations occur, has been shifted due to interactions with matter.  Translating our results into the infinite dimensional language, we would like to complete the square on the following expression.  Including the coupling constant $k$,

\begin{eqnarray}
\label{QUADR}
{1 \over {k}}\int_{\Sigma}\int_{\Sigma}{d^3x}{d^3y}
\biggl[{1 \over 2}A^a_{i}(\boldsymbol{x})\Gamma^{ij}_{ab}(\boldsymbol{x},\boldsymbol{y})
A^b_{j}(\boldsymbol{y})
+\Gamma^i_{a}(\boldsymbol{x})
\delta^{(3)}(\boldsymbol{x}-\boldsymbol{y})A^a_{i}(y)+k^{-1}\Gamma_{\Sigma}\nonumber\\
+{1 \over 2}
\phi^{\alpha}(\boldsymbol{x})
\Gamma_{\alpha\beta}(\boldsymbol{x},\boldsymbol{y})\phi^{\beta}(\boldsymbol{y})
+\Gamma_{\alpha}(\boldsymbol{x})
\delta^{(3)}(\boldsymbol{x}-\boldsymbol{y})\phi^{\alpha}(\boldsymbol{y})
+A^a_{i}(\boldsymbol{x})
\bigl(\Gamma_{\alpha}(\boldsymbol{x},\boldsymbol{y})\bigr)^i_{a}\phi^{\alpha}(\boldsymbol{y})\biggr]
\end{eqnarray}

\noindent
Going through the analogous steps as in the finite dimensional case, one can either compute the gravitational path integral first to obtain

\begin{eqnarray}
\label{QUADR1}
Z(0,0)=\int{DA}{D\phi}\Psi[A,\phi]\Bigl\vert_{Gauss}
=k^{-(\zeta_1+\zeta_2)}\Bigl(\hbox{Det}^{-1/2}\Delta^{ab}_{ij}\Bigr)
\Bigl(\hbox{Det}^{-1/2}\widetilde{\Delta}^{\alpha\beta}\Bigr)\nonumber\\
\hbox{exp}\Bigl[{1 \over {2k}}\int_{\Sigma}\int_{\Sigma}{d^3x}{d^3y}\Gamma^i_{a}(x)\Delta^{ab}_{ij}(x,y)\Gamma^j_{b}(y)\Bigr]
\hbox{exp}\Bigl[{1 \over {2k}}\int_{\Sigma}\int_{\Sigma}{d^3x}{d^3y}\widetilde{\Gamma}_{\alpha}(x)\widetilde{\Delta}^{\alpha\beta}(\boldsymbol{x},\boldsymbol{y})\widetilde{\Gamma}_{\beta}(y)\Bigr]
\end{eqnarray}

\noindent
where we have defined

\begin{eqnarray}
\label{QUADR2}
\widetilde{\Gamma}_{\alpha}(x)=\Gamma_{\alpha}(x)
-\int_{\Sigma}\int_{\Sigma}d^{3}{u}d^3{v}\Gamma^i_{a}(u)
\Delta^{ab}_{ij}(u,v)\bigl(\Gamma_{\alpha}(v,x)\bigr)^j_{b}
\end{eqnarray}

\begin{eqnarray}
\label{QUADR2}
\widetilde{\Delta}^{\alpha\beta}(x,y)=\Bigl(\Gamma_{\alpha\beta}(x,y)
-\int_{\Sigma}\int_{\Sigma}d^{3}{u}d^3{v}\bigl(\Gamma_{\alpha}\bigr)^i_{a}\Delta^{ab}_{ij}(u,v)\bigl(\Gamma_{\beta}\bigr)^j_{b}\Bigr)^{-1}
\end{eqnarray}

\noindent
and $\zeta_1$ and $\zeta_2$ are the zeta functions associated to the unmodified gravitational and modified matter propagators, respectively.  The interpretation of (\ref{QUADR1}) is that one may compute the gravitational portion first, and then compute the matter portion with a `correction' due to gravity to the one and two-point vertices.  Alternatively, one may integrate the matter portion first to obtain

\begin{eqnarray}
\label{QUADR3}
Z(0,0)=\int{DA}{D\phi}\Psi[A,\phi]\Bigl\vert_{Gauss}
=k^{-(\zeta^{\prime}_1+\zeta^{\prime}_2)}\Bigl(\hbox{Det}\Delta^{ab}_{ij}\Bigr)\Bigl(\hbox{Det}\widetilde{\Delta}^{\alpha\beta}\Bigr)\nonumber\\
\hbox{exp}\Bigl[{1 \over {2k}}\int\int{d^3x}{d^3y}\Gamma_{\alpha}(x)\Delta^{\alpha\beta}(x,y)\Gamma_{\beta}(y)\Bigr]
\hbox{exp}\Bigl[{1 \over {2k}}\int\int{d^3x}{d^3y}\widetilde{\Gamma}^i_{a}(x)\widetilde{\Delta}^{ab}_{ij}(x,y)\widetilde{\Gamma}^j_{b}(y)
\end{eqnarray}

\noindent
where we have defined

\begin{eqnarray}
\label{QUADR4}
\widetilde{\Gamma}^i_{a}(x)=\Gamma^i_{a}(x)-\int_{\Sigma}\int_{\Sigma}d^3{u}d^3{v}\Gamma_{\alpha}(u)\Delta^{\alpha\beta}(u,v)\bigl(\Gamma_{\beta}(v,x)\bigr)^i_{a}
\end{eqnarray}

\begin{eqnarray}
\label{QUADR5}
\widetilde{\Delta}^{ab}_{ij}(x,y)=\Bigl(\Gamma^{ij}_{ab}(x,y)
-\int_{\Sigma}\int_{\Sigma}d^3{u}d^3{v}\bigl(\Gamma_{\alpha}(x,u)\bigr)^i_{a}\Delta^{\alpha\beta}\bigl(\Gamma_{\beta}(y,v)\bigr)^j_{b}\Bigr)^{-1}
\end{eqnarray}

\noindent
whereupon (\ref{QUADR5}) corresponds to an unmodified matter propagator with a gravitational propagator modified due to the presence of matter.  The modification in either case is
due to the cross term arising from the mixed partials condition.  $\zeta^{\prime}_1$ and $\zeta^{\prime}_2$ respectively are the zeta functions associated with the propagators
modified in the opposite order.  From (\ref{QUADR1}) and (\ref{QUADR3}) one can write by inspection the source current representation of the wavefunction by the replacements

\begin{equation}
\label{QUADR6}
\Gamma_{\alpha}\longrightarrow\Gamma_{\alpha}
+\eta_{\alpha};~~\Gamma^i_{a}\longrightarrow\Gamma^i_{a}+J^i_{a},
\end{equation}

\subsection{Higher order interaction terms}
\par
\medskip
\indent 
The three-dimensional effective action forming the argument of the exponential in the wavefunction of the universe is given by

\begin{eqnarray}
\label{EFF}
\Gamma_{\Sigma}=\sum_{n=0}^{\infty}{{k^{n/2}} \over {n!}}\int{d}^3{x_1}...\int{d}^3{x_n}
\Bigl(\sum_{m=0}^{n}{n \choose m}
\bigl(\Gamma_{\alpha_{m+1}...\alpha_n}(x_1,...x_n)\bigr)^{i_1...i_m}_{a_1...a_m}\nonumber\\
A^{a_1}_{i_1}(x_1)...A^{a_m}_{i_m}(x_m)\phi^{\alpha_{m+1}}(x_{m+1})\phi^{\alpha_n}(x_n)\Bigr).
\end{eqnarray}

\noindent
This can also be seen in terms of a translation of the fields from their reference configuration as determined by the vertices, to their final configuration which is arbitrary.
The wave function of the universe containing gravity coupled to matter fields can be written in terms of the fields living on a final spatial three-surface in the form

\begin{equation}
\Psi[A,\phi]=\Psi[A(\Sigma_T),\phi(\Sigma_T)]=e^{I}
\end{equation}

\noindent
One does not need an explicit expression for the wavefunction for an arbitrary configuration in order to compute the norm of $\Psi_{GKod}$, since one can expand it about an arbitrary reference configuration

\begin{equation}
\Psi[\alpha+A,\varphi+\phi]=\hbox{exp}\Bigl[\int_{\Sigma}d^3{x}\Bigl(A^i_{a}(x){\delta \over {\delta{\alpha^a_{i}(x)}}}
+\phi^{\alpha}(x){\delta \over {\delta\varphi^{\alpha}(x)}}\Bigr)\Bigr]\Psi[\alpha,\varphi]
\end{equation}

\noindent
The factors of $k$ in (\ref{EFF}) arise due to rescaling the Ashtekar connection $A^a_{i}\rightarrow\sqrt{k}A^a_{i}$ and the matter variables 
$\phi^{\alpha}\rightarrow\sqrt{k}\phi^{\alpha}$, in anticipation of a loop expansion in powers of the coupling constant $\sqrt{k}$.  The coefficients of the expansion, which 
form the analogue of the 1PI vertices, are given by

\begin{equation}
\label{EFF1}
\Gamma_{\alpha_{m+1}...\alpha_n}(x_1,...x_m)
=\prod_{k=1}^{m}{\delta \over {\delta{A^{a_k}_{i_k}(x_k)}}}
\prod_{l=m+1}^{n}{\delta \over {\delta\phi^{\alpha_l}(x_l)}}\Gamma_{\Sigma}\biggl\vert_{A=\alpha,\phi=\varphi}.
\end{equation}

\noindent
Due to the CDJ Ansatz and the mixed partials condition, all vertices are uniquely determined from the CDJ matrix elements.\par
\indent
To calculate the norm of the wavefunction, we will need to be able to evaluate the partition function for the generalized Kodama state in analogy to the Chern-Simons partition function.  So let us expand the quantity

\begin{equation}
\label{EFF3}
\int{DA}{D\phi}e^{(\hbar{G})^{-1}\Gamma_{\Sigma}},
\end{equation}

\noindent
where $\Gamma_{\Sigma}$ is given by (\ref{EFF}).  The partition function \cite{EFF3} is given by

\begin{equation}
\label{EFF4}
e^{\hat{V}[\delta/\delta{J},\delta/\delta\eta]}Z(J,\eta)\biggl\vert_{J=\eta=0},
\end{equation}

\noindent
where $Z(J,\eta)$ is the Gaussian part and the `interaction' operator $\hat{V}$ is given by

\begin{eqnarray}
\label{EFF5}
\hat{V}[\delta/\delta{J},\delta/\delta\eta]
=\sum_{n=3}^{\infty}{1 \over {n!}}\Bigl(\prod_{k=1}^n\int_{\Sigma}d^{3}x_k\Bigr)k^{{n/2}-1}\nonumber\\
\biggl[\sum_{m=0}^{n}{n \choose m}
\bigl(\Gamma_{\alpha_{m+1}...\alpha_n}(x_1,...x_n)\bigr)^{i_1...i_m}_{a_1...a_m}
\prod_{k=1}^{m}\prod_{l=m+1}^{n}{\delta \over {\delta{J}^{i_k}_{a_k}(x_k)}}{\delta \over {\delta\eta_{\alpha_l}(x_l)}}\biggr]
\end{eqnarray}

\noindent
The exponential (\ref{EFF4}) can be written as an infinite product

\begin{equation}
\label{EFF6}
e^{\hat{V}}
=\hbox{exp}\Bigl(\sum_{n=3}^{\infty}k^{{n/2}-1}{{\hat{A}_{n}} \over {n!}}\Bigr)
=\prod_{n=3}^{\infty}\hbox{exp}\Bigl(k^{{n/2}-1}{{\hat{A}_{n}} \over {n!}}\Bigr)
\end{equation}

\noindent
where the operator $\hat{A}_n$ is given by

\begin{eqnarray}
\label{EFF7}
\hat{A}_{n}
=\sum_{m=0}^{n}\Lambda{n \choose m}\Bigl(\prod_{k=1}^n\int_{\Sigma}d^{3}x_k\Bigr)
\bigl(\Gamma_{\alpha_{m+1}...\alpha_n}(x_1,...x_n)\bigr)^{i_1...i_m}_{a_1...a_m}\nonumber\\
\prod_{k=1}^{m}\prod_{l=m+1}^{n}{\delta \over {\delta{J}^{i_k}_{a_k}(x_k)}}{\delta \over {\delta\eta_{\alpha_l}(x_l)}}
\end{eqnarray}

\noindent
Continuing with (\ref{EFF6}), we have

\begin{eqnarray}
\label{EFF8}
e^{\hat{V}}=\prod_{n=3}^{\infty}\hbox{exp}\Bigl(k^{{n/2}-1}{{\hat{A}_{n}} \over {n!}}\Bigr)
=\prod_{n=3}^{\infty}\sum_{m=0}^{\infty}{1 \over {m!}}k^{m({n/2}-1)}\Bigl({{\hat{A}_{n}} \over {n!}}\Bigr)^m\nonumber\\
=\sum_{n_1,n_2,...}\prod_{k}{1 \over {m_k!}}(1/{n_k}!)^{m_k}k^{m_{k}({n_{k}/2}-1)}(\hat{A}_{n_k})^{m_k}
\end{eqnarray}

\noindent
Note that in (\ref{EFF8}) we have interchanged the product and summation sequences.  The interpretation is that one takes the sum over all paths (histories) through an infinite dimensional lattice with coordinates labeled by $(m,n)$.  Here $n_k$ labels the point along a particular path and $m_k$ labels the value corresponding to that point.  This is the discrete analogue of the Feynman summation over histories in a field theory, in that all paths are continuous.

\section{Renormalizability of the norm}

\indent
The title of this section may seem counterintuitive, as one usually associates renormalizability with the path integral corresponding to the wavefunction or the transition amplitude of the
system.  In our case, only the criteria of finiteness applies to the transition amplitude as shown in \cite{EYOPATH}.  We will find that in order to compute the norm of $\Psi_{GKod}$ we will need to assess the issue of renormalizability.\par
\indent

\subsection{Perturbative approach to the generalized Kodama partition function}

Let us put the terms of the expansion in standard notation.  It may be helpful to combine the gravitational and matter fields, viewed as resulting from a Kaluza--Klein reduction, into their original form as a dimensionally extended system.  Denote unified quantities with upper case indices.  A generalized source current is given by 
$J^I_A\equiv(J^i_{a},\eta_{\alpha})$ and a generalized field is given by 
$\Phi\equiv(G^{-1}A^a_i,\phi^{\alpha})$ and a general n-point interaction vertex is given by

\begin{eqnarray}
\label{INTER}
\Gamma^{(N)}\Phi_{(N)}\sim\Gamma^{I_{1}...I_{N}}_{A_{1}...A_{N}}\Phi^{A_{1}}_{I_1}...\Phi^{A_{N}}_{I_N}
\end{eqnarray}

\noindent
where the appropriate index structure for contractions is in place in (\ref{INTER}).  The Gaussian part of the norm, including the source current $J^I_{A}$ is given by

\begin{eqnarray}
\label{INTER1}
Z(J)=(\hbox{Det}^{-1/2}\Gamma^{IJ}_{AB})e^{(\hbar{G})^{-1}\Gamma_0}
\hbox{exp}\Bigl[{k \over 2}\Delta^{AB}_{IJ}(\Gamma^I_A+J^I_A)(\Gamma^J_B+J^J_B)\Bigr],
\end{eqnarray}

\noindent
where $\Delta^{AB}_{IJ}\sim(\Gamma^{-1})^{AB}_{IJ}$ is a `generalized' propagator encompassing both gravitational and matter variables.\footnote{The functional determinant should in principle be computable via heat kernel methods, which we relegate to separate work.}  In accordance with (\ref{EFF8}) the
interaction terms are given by

\begin{eqnarray}
\label{INTER3}
Interaction~term
=e^{\hat{V}(\delta/\delta{J})}
=\sum_{N_{1},N_{2},...}^{\infty}k^{(1/2(N_{1}m_{1}+N_{2}m_{2}+...-(m_1+m_2+...))}\nonumber\\
{{(N_{1}!)^{-m_1}} \over {m_{1}!}}{{(N_{2}!)^{-m_2}} \over {m_{2}!}}...
\Gamma^{I_{a_1}...I_{a_{N_1}}}_{A_{a_1}...I_{A_{N_1}}}\Gamma^{I_{a_2}...I_{a_{N_2}}}_{A_{a_2}...I_{A_{N_2}}}...
{\delta \over {\delta{J}^{I_{a_1}}_{A_{a_1}}}}{\delta \over {\delta{J}^{I_{a_2}}_{A_{a_2}}}}...
\end{eqnarray}

\noindent
So all that remains is to evaluate

\begin{eqnarray}
\label{INTER4}
\vert\Psi\vert^{2}=\int{DA}D\phi\Psi_{GKod}^{*}[A,\phi]\Psi_{GKod}[A,\phi]=e^{\hat{V}(\delta/\delta{J})}Z(J)\Bigl\vert_{J=0}.
\end{eqnarray}

\noindent
Equation (\ref{INTER4}) generates a renormalizable series of Feynman diagrams, renormalizable due to the dimensionless coupling constant $k=\sqrt{\hbar{G}\Lambda}$.  As to whether the entire perturbative series is convergent we will save for a more detailed analysis of Borel summability.  It appears on first sight that the combination of the factorials in the denominator in (\ref{INTER3}) combined with
the high powers of $k$ should be more than sufficient to outweigh the symmetry factors due to the Feynman diagrammatic expansion.  So not only is the generalized Kodama $\Psi_{GKod}$ state normalizable, its partition function might (superficially) convergent.  This analysis, of course, is based on the existence of a solution for the CDJ matrix elements $\Psi_{ab}$, and as well requires further investigation to establish or to rule out.  We will provide a heuristic argument for convergence in the next section, where we compute the effective action nonperturbatively.  As an Ansatz to calculate the expectation value of operators $\hat{O}$ one can insert the operator 
into (\ref{INTER4}), as in

\begin{eqnarray}
\label{INTER5}
\bigl<\hat{O}\bigr>=\int{DA}D\phi\Psi_{GKod}^{*}(\hat{O}[A,\phi])[A,\phi]\Psi_{GKod}[A,\phi]
=e^{\hat{V}(\delta/\delta{J})}\hat{O}[\delta/\delta{J}]Z(J)\Bigl\vert_{J=0}.
\end{eqnarray}

\noindent
Note that in the case of a quadratic starting action $\Gamma$, $V=0$ and the result for the norm of the wavefunction reduces to the two-loop term.

\subsection{Alternate nonperturbative representation of the generalized Kodama partition function}

\noindent
We now transform the path integral into a new representation.  We will demonstrate with a generalized field $\phi=\phi(x)$ for simplicity, where the field can be generalized to include any number of components.  From the field one can define a wavefunction $\Psi$ such that

\begin{eqnarray}
\label{WAVEFUN}
\Psi[\phi]=e^{-{{\Gamma[\phi]} \over g}}.
\end{eqnarray}

\noindent
We now wish to evaluate the path integral of (\ref{WAVEFUN}).  By the methods of the previous sections we must choose a reference configuration $\varphi$ of the field $\phi$ and expand the path integral in fluctuations about this reference configuration.  The path integral then defines a new `effective' wavefunction $\Psi_{eff}[\varphi]$ based upon this configuration.

\begin{eqnarray}
\label{WAVEFUN1}
\Psi_{eff}[\varphi]=e^{-{{\Gamma[\varphi]} \over g}}
=\int{D\phi}e^{-{{\Gamma[\varphi+\phi]} \over g}}
=\int{D\phi}\Psi[\varphi+\phi]
\end{eqnarray}

\noindent
Since the reference configuration $\varphi$ can be chosen arbitrarily, one expects that the partition function should be independent of this configuration.  We will derive in this section a criterion for this indepedence.  By the standard procedure we expand the argument of the exponential in a functional Taylor series,

\begin{eqnarray}
\label{WAVEFUN2}
\Gamma[\varphi+\phi]=\Gamma[\varphi]
+\int_{\Sigma}d^3x{{\delta\Gamma} \over {\delta\varphi(x)}}\phi(x)
+{1 \over {2!}}\iint{d^3x}{d^3y}{{\delta^2\Gamma} \over {\delta\varphi(x)\delta\varphi(y)}}
\phi(x)\phi(y)+V[\varphi+\phi]
\end{eqnarray}

\noindent
where we have defined the interaction part $V$ by the expression

\begin{eqnarray}
\label{WAVEFUN3}
V[\varphi+\phi]=\sum_{n=3}^{\infty}{1 \over {n!}}
\int_{\Sigma}d^3x_1\dots\int_{\Sigma}d^3x_n
{{\delta^n\Gamma} \over {\delta\varphi(x_1)\dots\delta\varphi(x_n)}}
\phi(x_1)\dots\phi(x_n)
\end{eqnarray}

\noindent
Next, we complete the square by making the identification $J(x)=\delta\Gamma/\delta\varphi(x)$ and rescale the variable $\phi\rightarrow\sqrt{g}\phi$.  This sets the scale of quantum fluctuations to the dimensionless coupling constant $\sqrt{g}$.  By performing the standard integration of the Gaussian term we arrive at the standard relation

\begin{eqnarray}
\label{WAVEFUN4}
\Psi_{eff}[\varphi]=g^{\zeta(0)/2}e^{-\zeta^{\prime}(0)/2}
e^{-{{\Gamma[\varphi]} \over g}}
\hbox{exp}\Bigl[-{1 \over g}V[\varphi+\sqrt{g}{\delta \over {\delta{J}}}]\Bigr]\nonumber\\
\hbox{exp}\Bigl[{1 \over {2g}}\iint{d^3x}d^3yJ(x)\Delta(x,y)J(y)\Bigr]\biggl\vert_{J=0}
\end{eqnarray}

\noindent
As an aside, we note that the pre-factors in (\ref{WAVEFUN4}) contain information about the 
spatial manifold $\Sigma$.\par
\indent
We have expressed the path integral in terms of the standard perturbative scheme in which the interactions act on the propagator term to generate the loop expansion.  Our goal is now to transform (\ref{WAVEFUN4}) into a representation that encapsulates the nonperturbative effect of the entire series in compact form.  The rationale is that all of the information determining the Feynman diagrams is contained in the 1PI vertices in the potential $V$.  Since unlike in nonrenormalizable theories, the specific form of $V$ is known to all orders, there should be no need to introduce counterterms into the starting action $\Gamma$.\par
\indent
First, we must motivate the concept of a Fock space vacuum with respect to the 
configuration $\varphi$.  Note that we have defined the analogue of the propagator in three-space by

\begin{eqnarray}
\label{WAVEFUN5}
\Delta(x,y)=\Delta(\boldsymbol{x},\boldsymbol{y})
\sim\bigl<0\bigr\vert\varphi(x)\varphi(y)\bigl\vert{0}\bigr>_{\varphi}
=\Bigl({{\delta^2\Gamma} \over {\delta\varphi(x)\delta\varphi(y)}}\Bigr)^{-1}
\end{eqnarray}

\noindent
which serves as the inverse of the kinetic operator.  In general this is a curved space propagator, and we have assumed that it corresponds to a unique Fock space vacuum which enables an expansion of the field and its momentum in terms of mode creation and annihilation operators.\cite{ASHT}

\begin{eqnarray}
\label{WAVEFUN6}
\varphi(x)=\int{{d^3k} \over {\sqrt{2\omega(k)}}}
\bigl(a^{\dagger}(k)f_k(x)+ia(k)f_{\overline{k}}(x)\bigr)
\end{eqnarray}

\noindent
with corresponding conjugate momentum

\begin{eqnarray}
\label{WAVEFUN7}
\pi(x)=\int{{d^3p} \over {\sqrt{2\omega(p)}}}
\bigl(a^{\dagger}(p)f_p(x)-ia(p)f_{\overline{p}}(x)\bigr)
\end{eqnarray}

\noindent
where the basis functions are orthonormal with respect to the weight $\omega$.  Assuming commutation relations between the mode operators

\begin{eqnarray}
\label{WAVEFUN8}
[a(k),a(p)]=[a^{\dagger}(k),a^{\dagger}(p)]=0;~~
[a(k),a^{\dagger}(p)]=\delta^{(3)}(k,p)
\end{eqnarray}

\noindent
we see that the corresponding commutation relations for the field $\varphi$

\begin{eqnarray}
\label{WAVEFUN9}
\bigl[\hat{\varphi}(x),\hat{\varphi}(y)\bigr]
=\bigl[\hat\pi(x),\hat\pi(y)\bigr]=0;~~
\bigl[\hat{\varphi}(x),\hat\pi(y)\bigr]=i\delta^{(3)}(\boldsymbol{x},\boldsymbol{y})
\end{eqnarray}

\noindent
are consistent provided that the eigenfunctions obey the completeness relations

\begin{eqnarray}
\label{WAVEFUN10}
\int{{d^3k} \over {2\omega(k)}}f_k(x)f_{\overline{k}}(y)=\delta^{(3)}(\boldsymbol{x},\boldsymbol{y})
\end{eqnarray}

\par
\noindent
Next, we note that in the usual field theories about a Fock space vacuum, the following relation exists due to Wick's theorem\footnote{We have related the Fock space ground state to a generating function for correlation functions as in the usual quantum field theories on Minkowski spacetime.  The difference in this case is that the Fock space vacuum (which is uniquely taylored to the reference field configuration $\varphi$) is general is not the same as the Minkowski vacuum.}

\begin{eqnarray}
\label{WAVEFUN11}
\hbox{exp}\Bigl[{1 \over {2g}}\iint{d^3x}d^3yJ(x)\Delta(x,y)J(y)\Bigr]
=\bigl<0\bigr\vert\hbox{exp}\Bigl[{1 \over {\sqrt{g}}}
\int_{\Sigma}d^3x\hat{\varphi}(x)J(x)\Bigr]\bigl\vert{0}\bigr>_{\varphi}
\end{eqnarray}

\noindent
We now make use of an identity

\begin{eqnarray}
\label{WAVEFUN12}
e^{F[\delta/\delta{J}]}e^{G[J]}\biggl\vert_{J=0}
=e^{G[\delta/\delta\phi]}e^{F[\phi]}\biggr\vert_{\phi=0}.
\end{eqnarray}

\noindent
The combination of (\ref{WAVEFUN11}) and (\ref{WAVEFUN12}) in (\ref{WAVEFUN4}) yields

\begin{eqnarray}
\label{WAVEFUN13}
\Psi_{eff}[\varphi]=g^{\zeta(0)/2}e^{-\zeta^{\prime}(0)/2}
e^{-{{\Gamma[\varphi]} \over g}}
\hbox{exp}\Bigl[{1 \over {2g}}\iint{d^3x}d^3y{\delta \over {\delta\phi(x)}}
\Delta(x,y){\delta \over {\delta\phi(y)}}\Bigr]
\hbox{exp}\Bigl[-{1 \over g}V[\varphi+\sqrt{g}\phi]\Bigr]
\biggl\vert_{\phi=0}\nonumber\\
=g^{\zeta(0)/2}e^{-\zeta^{\prime}(0)/2}
e^{-{{\Gamma[\varphi]} \over g}}
\bigl<0\bigr\vert\hbox{exp}\Bigl[{1 \over {\sqrt{g}}}
\int_{\Sigma}d^3x\hat{\varphi}(x){\delta \over {\delta\phi(x)}}\Bigr]\bigl\vert{0}\bigr>_{\varphi}
\hbox{exp}\Bigl[-{1 \over g}V[\varphi+\sqrt{g}\phi]\Bigr]\biggl\vert_{\phi=0}.
\end{eqnarray}

\noindent
Since the $\varphi$ dependence in $V$ is in the form of c-numbers through the complete set of $1PI$ vertices, the last term in (\ref{WAVEFUN13}) can be brought within the ket ground state.  In this case the functional derivative acts on the $\phi$ dependence and `translates' the potential term by the reference configuration $-\varphi$.  This sequence of steps leads to

\begin{eqnarray}
\label{WAVEFUN14}
\Psi_{eff}[\varphi]=g^{\zeta(0)/2}e^{-\zeta^{\prime}(0)/2}
e^{-{{\Gamma[\varphi]} \over g}}
\bigl<0\bigr\vert\hbox{exp}\Bigl[{1 \over {\sqrt{g}}}
\int_{\Sigma}d^3x\hat{\varphi}(x){\delta \over {\delta\phi(x)}}\Bigr]
\hbox{exp}\Bigl[-{1 \over g}V[\varphi+\sqrt{g}\phi]\Bigr]\bigl\vert{0}\bigr>_{\varphi}
\biggl\vert_{\phi=0}\nonumber\\
=g^{\zeta(0)/2}e^{-\zeta^{\prime}(0)}e^{-{{\Gamma[\varphi]} \over g}}
\bigl<0\bigl\vert\hbox{exp}\Bigl[-{1 \over g}V[\varphi-\hat{\varphi}+\sqrt{g}\phi]\Bigr]
\bigl\vert{0}\bigr>\biggl\vert_{\phi=0}
\end{eqnarray}

\noindent
The aforementioned operation is tantamount to determining all of the connected Feynman diagrams in the partition function $Z=e^W[J]$.  Once this operation is performed, then one sets $J=0$, which is the anoalogue of setting $\phi=0$ in (\ref{WAVEFUN14}).  This leads to following compact form of the pathintegral in terms of the wavefunction

\begin{eqnarray}
\label{WAVEFUN15}
\Psi_{eff}[\varphi]=g^{\zeta(0)/2}e^{-\zeta^{\prime}(0)/2}
\Psi[\varphi]\bigl<0\bigr\vert{e}^{-{1 \over g}V[\varphi-\hat{\varphi}]}\bigl\vert{0}\bigr>_{\varphi}
\nonumber\\
\end{eqnarray}

\noindent
The net result is that the partition function is the wavefunction evaluated on a reference configuration times a wavefunction representing the fluctuations about that 
configuration.  Equation (\ref{WAVEFUN15}) can be written in the following form

\begin{eqnarray}
\label{WAVEFUN16}
\Psi_{eff}[\varphi]\propto
\Bigl<0\Bigr\vert{{\Psi_{GKod}[\varphi-\hat{\varphi}]} \over {\Psi_2}}\Bigl\vert{0}\Bigr>_{\varphi}
\end{eqnarray}

\noindent
where $\Psi_2$ corresponds to the Gaussian part of the wavefunction.

\section{Quantized Hamiltonian constraint in the language of 1PI vertices}

\indent
Let us rewrite the quantized Hamiltonian constraint introduced in \cite{EYO}in the language of 1PI vertices.  We will first write the ingredients prior to contraction with double epsilon tensors. 
In a regularized treatment in usual field theory one might take an expression of the form

\begin{eqnarray}
\label{REGULARIZED}
\hat{H}\Psi_{GKod}=
\hbox{lim}_{\epsilon\rightarrow{0}}f_{\epsilon}(x,y,z)\epsilon_{abc}\epsilon^{ijk}\Bigl[{\Lambda \over 6}\hbar^3{G}^3{\delta \over {\delta{A}^a_i(x)}}
{\delta \over {\delta{A}^b_j(y)}}{\delta \over {\delta{A}^c_k(z)}}\nonumber\\
+\hbar^2{G}^2{\delta \over {\delta{A}^a_i(x)}}{\delta \over {\delta{A}^b_j(y)}}B^k_{c}(z)+G\hat{\Omega}\Bigr]
e^{(G\hbar)^{-1}\Gamma_{\Sigma}}=0
\end{eqnarray}

\noindent
where the regulating function $f_{\epsilon}(x,y,z)$ imposes the necessary symmetry of the spatial positions $x$, $y$, $z$ and also enforces the coincidence limit 
$x\rightarrow{y}\rightarrow{z}$ in the limit $\epsilon\rightarrow{0}$.  Likewise for the matter contribution, taking a Klein--Gordon scalar field coupled to gravity in Ashtekar
variables,

\begin{eqnarray}
\label{MATTER}
G\hbox{lim}_{\epsilon\rightarrow{0}}f_{\epsilon}(x,y)\Bigl[\hbar^2{G^2}
T_{ij}{\delta \over {\delta{A}^a_i(x)}}{\delta \over {\delta{A}^a_j(y)}}
-{1 \over 2}\hbar^2{\delta \over {\delta\phi(x)}}{\delta \over {\delta\phi(y)}}\Bigr]e^{(G\hbar)^{-1}\Gamma_{\Sigma}}
\end{eqnarray}

\noindent
with its regulating function $f_{\epsilon}(x,y)$ imposing the necessary symmetry of the spatial positions $x$, $y$ and also enforcing the corresponding coincidence limit 
$x\rightarrow{y}$ in the limit $\epsilon\rightarrow{0}$.\par
\indent
However, we do not perform regularization on the canonical part of the quantization procedure.  We will suppress all dependence upon position in the following 
computation.  We start with the basic ingredients of the cosmological contribution, at the level prior to contraction with double epsilon tensors.

\begin{eqnarray}
\label{COSMOS}
\hbar^3{G}^3{\delta \over {\delta{A}^a_i}}{\delta \over {\delta{A}^b_j}}{\delta \over {\delta{A}^c_k}}e^{(G\hbar)^{-1}\Gamma_{\Sigma}}\nonumber\\
=\hbar^3{G}^3{\delta \over {\delta{A}^a_i}}{\delta \over {\delta{A}^b_j}}\bigl((G\hbar)^{-1}\Gamma^k_c\bigr)e^{(G\hbar)^{-1}\Gamma_{\Sigma}}
=\hbar^2{G}^2{\delta \over {\delta{A}^a_i}}{\delta \over {\delta{A}^b_j}}\Gamma^k_{c}e^{(G\hbar)^{-1}\Gamma_{\Sigma}}\nonumber\\
=\hbar^2{G}^2{\delta \over {\delta{A}^a_i}}\Bigl(\Gamma^{kj}_{cb}+(G\hbar)^{-1}\Gamma^k_{c}\Gamma^j_{b}\Bigr)e^{(G\hbar)^{-1}\Gamma_{\Sigma}}\nonumber\\
=\hbar{G}\Bigl(\hbar{G}\Gamma^{kji}_{cba}+\Gamma^{ki}_{ca}\Gamma^j_{b}+\Gamma^{ji}_{ba}\Gamma^k_{c}+\Gamma^{kj}_{cb}\Gamma^i_{a}
+(G\hbar)^{-1}\Gamma^k_{c}\Gamma^j_{b}\Gamma^i_{a}\Bigr)e^{(G\hbar)^{-1}\Gamma_{\Sigma}}\nonumber\\
=\Bigl(\hbar^2{G}^2\Gamma^{kji}_{cba}
+\hbar{G}\bigl(\Gamma^{ki}_{ca}\Gamma^j_{b}+\Gamma^{ji}_{ba}\Gamma^k_{c}+\Gamma^{kj}_{cb}\Gamma^i_{a}\bigr)
+\Gamma^k_{c}\Gamma^j_{b}\Gamma^i_{a}\Bigr)e^{(G\hbar)^{-1}\Gamma_{\Sigma}}
\end{eqnarray}

\noindent
We now move on the the curvature contribution

\begin{eqnarray}
\label{CURVA}
\hbar^2{G}^2{\delta \over {\delta{A}^a_i}}{\delta \over {\delta{A}^b_j}}B^k_{c}e^{(G\hbar)^{-1}\Gamma_{\Sigma}}\nonumber\\
=\hbar^2{G^2}{\delta \over {\delta{A}^a_i}}\bigl(D^{kj}_{cb}+(G\hbar)^{-1}B^k_{c}\Gamma^j_{b}\bigr)e^{(G\hbar)^{-1}\Gamma_{\Sigma}}
=\hbar{G}{\delta \over {\delta{A}^a_i}}\bigl(\hbar{G}D^{kj}_{cb}+B^k_{c}\Gamma^j_{b}\bigr)e^{(G\hbar)^{-1}\Gamma_{\Sigma}}\nonumber\\
=\Bigl(\hbar^2{G}^2\epsilon^{kji}_{cba}+\hbar{G}\bigl(D^{ki}_{ca}\Gamma^j_{b}+B^k_{c}\Gamma^{ji}_{ba}
+D^{kj}_{cb}\Gamma^i_{a}\bigr)+B^k_{c}\Gamma^j_{b}\Gamma^i_{a}\Bigr)e^{(G\hbar)^{-1}\Gamma_{\Sigma}}.
\end{eqnarray}

\noindent
The SQC would be identically satisfied by grouping coefficients of the quantum terms in (\ref{COSMOS}) and (\ref{CURVA}), if not for the matter contribution, giving as an
expansion in powers of $\hbar{G}$ of the eigenvalue of the constraint,

\begin{eqnarray}
\label{COMB}
\epsilon_{ijk}\epsilon^{abc}\Bigl[
\Gamma^j_{b}\Gamma^i_{a}\bigl(B^k_{c}+{\Lambda \over 6}\Gamma^k_{c}\bigr)
+\hbar{G}\Bigl(\Gamma^j_{b}\bigl(\Gamma^{ki}_{ca}+D^{ki}_{ca}\bigr)
+\Gamma^i_{a}\bigl(D^{kj}_{cb}+{\Lambda \over 6}\Gamma^{kj}_{cb}\bigr)\nonumber\\
+\Gamma^{ji}_{ba}\bigl(B^k_{c}+{\Lambda \over 6}\Gamma^k_{c}\bigr)\Bigr)
+\hbar^2{G^2}\bigl(\epsilon^{kji}_{cba}+{\Lambda \over 6}\Gamma^{kji}_{cba}\bigr)\Bigr]=0
\end{eqnarray}

\noindent
The self-duality condition is the requirement that the individual terms in brackets in 
(\ref{COMB}) cancel in pairs.  This imposes conditions from quantum gravity upon the 
Feynman vertices that the SQC must hold upon quantization of the theory.  Let us compute the contribution to this relation when a Klein-Gordon scalar field is introduced into the
theory.

\begin{eqnarray}
\label{MATTER1}
\Bigl[\hbar^2{G^2}T_{ij}{\delta \over {\delta{A}^a_i}}{\delta \over {\delta{A}^a_j}}
-{1 \over 2}\hbar^2{\delta \over {\delta\phi}}{\delta \over {\delta\phi}}\Bigr]e^{(G\hbar)^{-1}\Gamma_{\Sigma}}\nonumber\\
=\Bigl[\hbar{G}T_{ij}{\delta \over {\delta{A}^a_i}}(\Gamma^j_a)
-{{\hbar} \over {2G}}{\delta \over {\delta\phi}}(\Gamma_{\phi})\Bigr]e^{(G\hbar)^{-1}\Gamma_{\Sigma}}\nonumber\\
=\Bigl(T_{ij}\Gamma^i_{a}\Gamma^j_{a}+{1 \over 2}G^{-2}\Gamma_{\phi}\Gamma_{\phi}
+\hbar{G}\bigl(T_{ij}\Gamma^{ji}_{aa}-{i \over 2}G^{-2}\Gamma_{\phi\phi}\bigr)\Bigr)e^{(G\hbar)^{-1}\Gamma_{\Sigma}}
\end{eqnarray}

\noindent
By grouping coefficients of powers of $\hbar{G}$ we see that the vertex relations are modified, upon contraction of all tensors, into

\begin{eqnarray}
\label{COMB3}
T_{ij}\Gamma^i_{a}\Gamma^j_{a}+{1 \over 2}G^{-2}\Gamma_{\phi}\Gamma_{\phi}+\epsilon_{ijk}\epsilon^{abc}\Gamma^j_{b}\Gamma^i_{a}\bigl(B^k_{c}+{\Lambda \over 6}\Gamma^k_{c}\bigr)=0
\end{eqnarray}

\noindent
which corresponds to the condition that the semiclassical part of the Hamiltonian constraint vanishes $q_{0}=0$,

\begin{eqnarray}
\label{COMB4}
T_{ij}\Gamma^{ji}_{aa}-{i \over 2}G^{-2}\Gamma_{\phi\phi}\nonumber\\
\epsilon_{ijk}\epsilon^{abc}\Bigl(\Gamma^j_{b}\bigl(\Gamma^{ki}_{ca}+D^{ki}_{ca}\bigr)
+\Gamma^i_{a}\bigl(D^{kj}_{cb}+{\Lambda \over 6}\Gamma^{kj}_{cb}\bigr)
+\Gamma^{ji}_{ba}\bigl(B^k_{c}+{\Lambda \over 6}\Gamma^k_{c}\bigr)\Bigr)=0
\end{eqnarray}

\noindent
which corresponds to the condition that the part first-order in singularity vanishes 
($q_{1}=0$), and

\begin{eqnarray}
\label{COMB5}
36+{\Lambda \over 6}\epsilon_{ijk}\epsilon^{abc}\Gamma^{kji}_{cba}=0
\end{eqnarray}

\noindent
which is the condition that the second-order part vanishes ($q_{2}=0$).  We can see from 
(\ref{COMB3}), (\ref{COMB4}) and (\ref{COMB5}) that different matter models will impose
different relations amongst the vertices of the wavefunction of the universe and consequently upon its norm.

\section{Discussion}

We have provided a prescription for computing the partition function of the generalized Kodama states by importing techniques from Chern--Simons perturbation theory.  There is reason to infer normalizability of the loop expansion for this partition function based upon two main considerations.  First, the effective action for the generalized Kodama partition function is uniquely fixed by the first-order vertex of the tree-level action which is not true for nonrenormalizable theories, which require an infinite number of parameters to describe the quantum theory.  This is a direct consequence of the CDJ Ansatz and the mixed partials condition \cite{EYO}, imposed due to the semiclassical-quantum correspondence as a requirement for a finite wavefunction.  Secondly, the partition 
function for $\Psi_{GKod}$ features the same dimensionless coupling constant as its counterpart for the pure Kodama state $\Psi_{Kod}$, the latter of which is clearly renormalizable as well as finite.  We have also provided a geometric interpretation for some of the structures appearing in the 1PI vertices, highlighting the analogy to an infinite dimensional version of Kaluza-Klein theory applied to the functional space of fields.  The existence of the partition function for $\Psi_{GKod}$ is directly linked to the existence of $\Psi_{GKod}$ itself, which is a requirement imposed by the semiclassical-quantum correspondence for a finite state of quantum gravity.  Once examples of these states have been provided, then a future direction of research will include extending the prescription for computing this partition function to evaluation of the norm of the generalized Kodama states.  We have provided a naive Ansatz in the present paper for such a norm and expectation values, which should be examined in further detail to make a definite conclusion.  Nevertheless, the ability to define the partition function has definite applications irrespective of the norm of the generalized Kodama states.  One potential application, which we also present as a further line of research, is to transform $\Psi_{GKod}$ into the loop and spin network representations by the method of source currents as developed in the present work.  The loop state is then given by

\begin{eqnarray}
\label{LOOP}
\Psi[\gamma_1,\gamma_2,\dots,\gamma_n]=\int{DA}\Psi_{GKod}[A,\phi]\prod_{i=1}^ne^{\int_{\gamma_i}A}
\end{eqnarray}

\noindent
where the argument of the exponential of the Wilson lines is given, for a given curve $\gamma$ by

\begin{eqnarray}
\label{LOOP1}
\int_{\gamma}ds\dot{x}^iA^a_i(x(s))\tau_a
\end{eqnarray}

\noindent
In (\ref{LOOP1}) the source current $J^i_a$ can be chosen to correspond to the tangent vector of a network edge in the appropriate representation of the gauge group, parametrized by a curve $s$ in three space $\Sigma$.  Since the generalized Kodama state $\Psi_{GKod}$ already includes the effects of matter in a finite state exactly solving the constraints of general relativity, which as argued in \cite{EYO} posseses a direct link to the limit below the Planck scale, then one may be able to examine the manifestation and implications of such a state in the loop and spin network representations.\footnote{Equation (\ref{LOOP}) can be seen as the direct analogue of Jones polynomials discovered by Witten, in which $\Psi_{GKod}$ is replaced by $\Psi_{Kod}$, which incorporate the effects of matter fields quantized with gravity on the same footing.  This line of reasoning has been motivated by the development of Ed Witten in using the Chern--Simons partition function to determine knot and link invariants relevant to the study of quantum gravity.  We argue that the same developments should be applicable to the generalized Kodama states by direct analogy, with the beniefit of incorporating a well-defined semiclassical limit into the loop representation and examining its implications.}  Some outstanding issues to address, besides reality conditions for Ashtekar variables, is the rigorous proof of the existence of field configurations in the Ashtekar variables which guarantee converenge of the Gaussian part of the norm.\footnote{This is related to the existence of the appropriate contours of integration required.}

\section{Appendix: Toy model for Gaussian part of the norm}

Our method to compute the norm will involve an asymptotic expansion about the Gaussian part, which can be computed exactly.  Let us compute the Gaussian part first, making use 
of the shorthand notation for simplicity.

\begin{equation}
\label{ALG}
S_{2}(x,y)={1 \over 2}x^i(a_1)_{ij}x^j+(b_1)_{i}x^i+C+{1 \over 2}y^i(a_{2})_{ij}y^j+(b_2)_{i}y^i+x^{i}e_{ij}y^{j}
\end{equation}

\noindent
We first complete the square on $a_1$ in (\ref{ALG}), bringing in the matter contribution along with the linear term.  We will omit the indices for simplicity and reinsert them in
the full infinite dimensional case when completed.  We can make the following identifications when visualizing the infinite dimensional case

\begin{eqnarray}
\label{IDENTIFIC}
x\sim{A^a_i}(x);~~y\sim\phi^{\alpha}(x)\nonumber\\
a_1\sim\Gamma^{ij}_{ab}(x,y)\Bigl\vert_{A=a_{ref},\phi=\phi_{ref}};~~a_{2}\sim\Gamma_{\alpha\beta}(x,y)\Bigl\vert_{A=a_{ref},\phi=\phi_{ref}}\nonumber\\
b_1\sim\Gamma^i_{a}(x)\Bigl\vert_{A=a_{ref},\phi=\phi_{ref}};~~b_{2}\sim\Gamma_{\alpha}(x)\Bigl\vert_{A=a_{ref},\phi=\phi_{ref}}\nonumber\\
e\sim(\Gamma_{\alpha}(x,y))^i_a\Bigl\vert_{A=a_{ref},\phi=\phi_{ref}}\nonumber\\
C\sim\Gamma_{\Sigma}\Bigl\vert_{A=a_{ref},\phi=\phi_{ref}}
\end{eqnarray}

\noindent
Proceeding with the completion of the square,

\begin{eqnarray}
\label{ALG1}
{1 \over 2}a_{1}x^2+b_{1}x+exy+C+{1 \over 2}a_{2}y^{2}+b_{2}y
={1 \over 2}a_{1}x^2+(b_{1}+ey)x+C+{1 \over 2}a_{2}y^2+b_{2}y\nonumber\\
={1 \over 2}a_{1}\Bigl[x^2+{2 \over {a_1}}(b_{1}+ey)x+\Bigl({{b_{1}+ey} \over {a_1}}\Bigr)^2-\Bigl({{b_{1}+ey} \over {a_1}}\Bigr)^2\Bigr]
+C+{1 \over 2}a_{2}y^2+b_{2}y
\end{eqnarray}

\noindent
Thus the final result of completing the square for the Ashtekar connection $A^a_i$ is given by

\begin{equation}
\label{ALG2}
{1 \over 2}a_{1}\bigl(x+(a_1)^{-1}(b_1+ey)\bigr)^{2}
+C-{{(b_1+ey)^2} \over {2a_1}}+{1 \over 2}a_{2}y^2+b_{2}y
\end{equation}

\noindent
We must first evaluate the Gaussian integral over $x$ in (\ref{ALG2}).  This is given, upon rescaling the variable $x\rightarrow\sqrt{k}x$, by

\begin{equation}
\label{GAUSSIAN1}
\int{dx}~\hbox{exp}\Bigl[-{1 \over {2k}}{a_1}\bigl(x+(a_1)^{-1}(b_1+ey)\bigr)^{2}\Bigr]=\sqrt{{k \over a}} 
\end{equation}

\noindent
The condition for the Gaussian integral (\ref{GAUSSIAN1}) to converge is that $a_1>0$.  This places a constraint, in the infinite dimensional analogue, upon the reference field 
configuration for the Ashtekar variables which we are allowed to expand around of the form.  

\begin{eqnarray}
\label{INDENTIFIC1}
\Gamma^{ij}_{ab}(x,y)\Bigl\vert_{A=a_{ref},\phi=\phi_{ref}}~positive~definite
\end{eqnarray}

Note that the presence of the dynamical variable $y$ in (\ref{GAUSSIAN1}) corresponds to a shift
in $x$ space of the measure and does not affect the value of the integral.  In the language of field theory, the connection $A^a_i(x)$ and the matter field $\phi^{\alpha}(x)$ are
independent variables.\par
\indent
Once the first Gaussian is completed one can now expand (\ref{ALG2}) and complete the square with respect to $y$.  So we have

\begin{eqnarray}
\label{ALG3}
C-{{(b_1+ey)^2} \over {2a_1}}+{1 \over 2}a_{2}y^2+b_{2}y\nonumber\\
=C-{1 \over {2a_1}}\bigl(b_1^2+2b_1ey+e^2{y^2}\bigr)+{1 \over 2}a_{2}y^2+b_{2}y\nonumber\\
=C-{{b_1^2} \over {2a_1}}+{1 \over 2}\Bigl(a_{2}-{{e^2} \over {a_1}}\Bigr)y^{2}+\Bigl(b_{2}-{{b_{1}e} \over {a_1}}\Bigr)y\nonumber\\
=C-{{b_1^2} \over {2a_1}}+{1 \over 2}\Bigl(a_{2}-{{e^2} \over {a_1}}\Bigr)
\biggl[\Bigl(y+(a_2-{{e^2} \over {a_1}})^{-1}\bigl(b_{2}-{{b_{1}e} \over {a_1}}\bigr)\Bigr)^2\biggr]\nonumber\\
-{1 \over 2}\Bigl(a_2-{{e^2} \over {a_1}}\Bigr)^{-1}\Bigl(b_{2}-{{b_{1}e} \over {a_1}}\Bigr)^2                 
\end{eqnarray}

\noindent
One can now separate the `zeroth' order term from (\ref{ALG3})

\begin{equation}
\label{ZEROTH}
\Gamma_0\sim{C}-{{b_1^2} \over {2a_1}}-{1 \over 2}\Bigl(a_2-{{e^2} \over {a_1}}\Bigr)^{-1}\Bigl(b_{2}-{{b_{1}e} \over {a_1}}\Bigr)^2, 
\end{equation}

\noindent
and one is left with a Gaussian integral to perform over $y$, corresponding to the matter variables.  Rescaling the variable $y\rightarrow\sqrt{k}y$ to shift the coupling constant
to the interaction vertices, one has

\begin{equation}
\label{GAUSSIAN2}
\int{dy}~\hbox{exp}\Bigl[-{1 \over {2k}}\Bigl(a_2-{{e^2} \over {a_1}}\Bigr)
\Bigl(y+(a_2-{{e^2} \over {a_1}})^{-1}\bigl(b_{2}-{{b_{1}e} \over {a_1}}\bigr)\Bigr)^{2}\Bigr]
=\sqrt{k}(a_2-{{e^2} \over {a_1}})^{-1/2}
\end{equation}

\noindent
Again, the integral (\ref{GAUSSIAN2}) is invariant under the shift, which in addition to the first Gaussian, yields the condition that the Gaussian integral over the matter fields
converges for configurations for which

\begin{equation}
\label{CONDI}
a_{1}>0;~~a_2-{{e^2} \over {a_1}}>0
\end{equation}

\noindent
Equation(\ref{CONDI}) is equivalent in infinite dimensional language that in order to have a chance of a finite norm for the generalized Kodama states, we must have in addition to
(\ref{INDENTIFIC1}) that

\begin{eqnarray}
\label{INDENTIFIC2}
\Gamma_{\alpha\beta}(x,y)-\int_{\Sigma}\int_{\Sigma}{d}^3{x^{\prime}}d^3{y^{\prime}}\nonumber\\
(\Gamma_{\alpha}(x,x^{\prime})^i_{a}(\Gamma^{-1})^{ab}_{ij}(x^{\prime},y^{\prime})(\Gamma_{\beta}(y,y^{\prime}))^j_{b}\Bigl\vert_{A=a_{ref},\phi=\phi_{ref}}\nonumber\\
~positive~definite
\end{eqnarray}

\noindent
(\ref{INDENTIFIC2}) in the infinite dimensional language of field theory, imposes a necessary condition that the generalized Kodama state $\Psi_{GKod}$ not be unnormalizable.  This is a necessary condition.  It seems that it should in principle be possible to meet this, although its sufficiency requires deeper analysis.  Nevertheless, let us assume that it is possible to choose the reference configuration of the fields appropriately.\par
\indent
We can now evaluate the Gaussian part of the norm corresponding in condensed notation to 
(\ref{ALG}).  It is given by

\begin{eqnarray}
\label{GAUSSIAN3}
\int{Dx}{Dy}\hbox{exp}\bigl[-S_{2}(x,y)\bigr]\nonumber\\
=\bigl(\hbox{det}^{-1/2}a_1\bigr)\hbox{det}^{-1/2}\Bigl(a_2-{{e^2} \over {a_1}}\Bigr)
\hbox{exp}\Bigl[-C+{1 \over 2}(a_1)^{-1}_{ij}(b_1)^{i}(b_1)^{j}\nonumber\\
+{1 \over 2}\bigl((a_2)_{ij}-(a_1)_{kl}e^{ik}e^{jl}\bigr)^{-1}\bigl((b_2)^i-(a_1)^{-1}_{kl}(b_{1})_{k}e^{il}\bigr)
\bigl((b_2)^j-(a_1)^{-1}_{mn}(b_{1})_{m}e^{jn}\bigr)\Bigr]
\end{eqnarray}


\begin{thebibliography}{99}

\bibitem{EYO} {Eyo Eyo Ita III `Finite states in four dimensional quantized gravity' Class. Quantum Grav. 25 (2008) 125001}

\bibitem{PERT} {Justin Sawon `Perturbative expansion of Chern--Simons theory'
Geometry and Topology Monographs 8 (2006) 145-166}

\bibitem{PERT1} {David Adams `The semiclassical approximation for the Chern--Simons partition function'
arXiV:hep-th/9709147}

\bibitem{PERT3} {Dror Bar Nataan and Ed Witten `Perturbative expansion of Chern--Simons theory with non-compact gauge group'
Commun. Math. Phys. 141, 423-440 (1991)}

\bibitem{CSPT} {Chopin Soo, `Wavefunction of the universe and Chern-Simons perturbation theory'
Class. Quantum Grav. 19(2002)1051-1963}

\bibitem{BACKGROUND} {A. Rakhimov and Jae Hyung Yee `Optimized post Gaussian approximation in the background field method' 
arXiV:hep-th/0206121v3 11 August 2003}

\bibitem{JACKIW} {Roman Jackiw `Functional evaluation of the effective potential'
Phys. Rev. D Volume 9, Number 6 (1974)}

\bibitem{SCHROREP} {Richard P. Treat `Schr\"odinger representation of non-Abelian quantum gauge theory'
Phys. Rev. D14 Number 6 pp.1568-1576 (1976)}

\bibitem{WHEELER} {Misner, Thorne and Wheeler, `Gravitation' Copyright 1970 and 1971 by Misner, Thorne and Wheeler: 1973 by Freeman and Company}

\bibitem{AVRA2} {I.G. Avramidi `The covariant technique for calculation of the heat kernel asymptotic expansion'
Phys. Lett. B238 (1990)92-97}

\bibitem{AVRA3} {I.G. Avramidi `A new algebraic approach for calculating the heat kernel in gauge theories'
Phys. Lett. B305 (1993)27-34}

\bibitem{AVRA4} {The nonlocal structure of one-loop effective action via partial summation of asymptotic expansion'
Phys. Lett. B236 (1990)443}

\bibitem{PERT2} {David Adams `A note of the Fadeev--Poppov determinant and Chern--Simons perturbation theory'
ArXiV:hep-th/9704159}

\bibitem{ABELDUAL} {Emil M. Prodanov and Siddhartha Sen `Abelian duality'
Phys. Rev. D62, 045009}

\bibitem{WAVE3} {Halliwell and Hartle `Wave functions constructed from an invariant sum over histories satisfy constraints'
Phys. Rev. D43 (1991)1170}

\bibitem{EYOPATH} {Eyo Ita `Finite states in 4 dimensional quantized gravity. A brief introduction into the path integration approach in Ashtekar variables' 
arXiV:0804.0793[gr-qc]}

\bibitem{EYORENORM} {Eyo Eyo Ita `A brief survey of the renormalizability of four dimensional gravity for generalized Kodama states'
arXiV:gr-qc/0805.1893}

\bibitem{GEOMETRY} {G. Kunstatter `The path integral for gauge theories: a geometrical approach'
Class. Quantum Grav. 9 (1992) 157-168}

\bibitem{EYOFULL} {Eyo Ita `A systematic approach to the solution of the constraints of quantum gravity: The full theory.'}

\bibitem{ASHT} {Abhay Ashtekar and Anne Magnon `Quantum fields in curved spacetime'
Proceedings of the Royal Society of London. Series A, Mathematical and Physical Sciences, 
Vol.346, No. 1646 (1975) pp375-394}





















\end{thebibliography}
\end{document}